\documentclass[usegraphicx,usenatbib,letterpaper,twocolappendix]{emulateapj}

\usepackage{ulem}
\usepackage{color}
\usepackage{graphics,graphicx}
\usepackage{times} 
\usepackage{amssymb}
\usepackage{amsmath}
\usepackage{natbib}
\usepackage{url}
\usepackage{multirow}
\usepackage{ctable}
\usepackage{threeparttable}
\newif\ifAMStwofonts
\AMStwofontstrue

\def\xmm{{\it XMM-Newton}}

\def\rxte{{\it RXTE}}

\def\suzaku{{\it Suzaku}}

\def\chandra{{\it Chandra}}
\def\swift{{\it Swift}}
\def\integral{{\it INTEGRAL}}

\def\epicmos1{{EPIC-MOS1}}
\def\epicmos2{{EPIC-MOS2}}
\def\epicmos{{EPIC-MOS}}

\def\maxi{{\it MAXI}}
\def\nustar{{\it NuSTAR}}

\def\hitomi{{\it Hitomi}}

\def\deg{$^{\circ}$}

\def\rsun{\hbox{$\rm R_{\odot}$}}

\def\kmps{\hbox{km~s$^{-1}$}}

\def\H0{{\rm ~km~s^{-1}~Mpc^{-1}}}

\def\kev{\hbox{\rm keV}}

\def\ergpcmsqps{\hbox{$\rm\thinspace erg~cm^{-2}~s^{-1}$}}

\def\ergcmps{\hbox{\rm erg~cm~s$^{-1}$}}

\def\msun{\hbox{$\rm M_{\odot}$}}

\def\chisq{{$\chi^{2}$}}

\def\xspec{\hbox{\small XSPEC}}
\def\xspecv{\hbox{\small XSPEC}\, v12.6.0f}

\def\heasoft{\hbox{\rm{\small HEASOFT}}}
\def\nustardas{\rm {\small NUSTARDAS}}

\def\addascaspec{\hbox{\rm{\small ADDASCASPEC~\/}}}

\def\addascaspec{\hbox{\rm{\small ADDASCASPEC}}}

\def\xstar{\hbox{\rm{\small XSTAR}}}
\def\grid25{\hbox{\rm{\small GRID25}}}

\def\cutoffpl{\rm{\small CUTOFFPL}}

\def\tbabs{\rm{\small TBABS}}

\def\diskbb{\rm{\small DISKBB}}

\def\reflionx{\rm{\small REFLIONX}}
\def\xillver{\rm{\small XILLVER}}

\def\relconv{\rm{\small RELCONV}}
\def\relline{\rm{\small RELLINE}}

\def\gauss{\rm{\small GAUSSIAN}}

\def\fexxv{\hbox{\rm Fe\,{\small XXV}}}
\def\fexxvi{\hbox{\rm Fe\,{\small XXVI}}}

\def\ka{K\,$\alpha$}

\def\eg{{\it e.g.~\/}}

\def\ie{{\it i.e.~\/}}

\def\la{\mathrel{\hbox{\rlap{\hbox{\lower4pt\hbox{$\sim$}}}{\raise2pt\hbox{$<$}}}}}
\def\ga{\mathrel{\hbox{\rlap{\hbox{\lower4pt\hbox{$\sim$}}}{\raise2pt\hbox{$>$}}}}}

\def\d25{D$_{25}$}
\def\nh{{$N_{\rm H}$}}

\def\.25{0.25 keV\thinspace}

\def\rg{$R_{\rm{G}}$}

\def\rbr{$R_{\rm br}$}

\def\qin{{$q_{\rm{in}}$}}
\def\qout{{$q_{\rm{out}}$}}

\def\cyg{\rm Cygnus X-1}

\shorttitle{The Variable Corona of \cyg}
\shortauthors{D.~J. Walton et al.}

\begin{document}

\title{The Soft State of Cygnus X-1 Observed with \textit{NuSTAR}: A Variable Corona and a Stable Inner Disk}

\author{D. J. Walton\altaffilmark{1,2},
J. A. Tomsick\altaffilmark{3},
K. K. Madsen\altaffilmark{2},
V. Grinberg\altaffilmark{4},
D. Barret\altaffilmark{5,6},
S. E. Boggs\altaffilmark{3},
F. E. Christensen\altaffilmark{7},
M. Clavel\altaffilmark{3},
W. W. Craig\altaffilmark{3},
A. C. Fabian\altaffilmark{8},
F. Fuerst\altaffilmark{2},
C. J. Hailey\altaffilmark{9},
F. A. Harrison\altaffilmark{2},
J. M. Miller\altaffilmark{10},
M. L. Parker\altaffilmark{8},
F. Rahoui\altaffilmark{11,12},
D. Stern\altaffilmark{1},
L. Tao\altaffilmark{2},
J. Wilms\altaffilmark{13},
W. Zhang\altaffilmark{14},
}
\affil{
$^{1}$ Jet Propulsion Laboratory, California Institute of Technology, Pasadena, CA 91109, USA \\
$^{2}$ Space Radiation Laboratory, California Institute of Technology, Pasadena, CA 91125, USA \\
$^{3}$ Space Sciences Laboratory, University of California, Berkeley, CA 94720, USA \\
$^{4}$ MIT Kavli Institute for Astrophysics and Space Research, MIT, 70 Vassar Street, Cambridge, MA 02139, USA \\
$^{5}$ Universite de Toulouse, UPS-OMP, IRAP, Toulouse, France \\
$^{6}$ CNRS, IRAP, 9 Av. colonel Roche, BP 44346, F-31028 Toulouse cedex 4, France \\
$^{7}$ DTU Space, National Space Institute, Technical University of Denmark, Elektrovej 327, DK-2800 Lyngby, Denmark \\
$^{8}$ Institute of Astronomy, University of Cambridge, Madingley Road, Cambridge CB3 0HA, UK \\
$^{9}$ Columbia Astrophysics Laboratory, Columbia University, New York, NY 10027, USA \\
$^{10}$ Department of Astronomy, University of Michigan, 1085 S. University Ave, Ann Arbor, MI 48109-1107, USA \\
$^{11}$ European Southern Observatory, K. Schwarzschild-Str. 2, 85748 Garching bei München, Germany \\
$^{12}$ Department of Astronomy, Harvard University, 60 Garden street, Cambridge, MA 02138, USA \\
$^{13}$ ECAP-Erlangen Centre for Astroparticle Physics, Sternwartstrasse 7, D-96049 Bamberg, Germany \\
$^{14}$ NASA Goddard Space Flight Center, Greenbelt, MD 20771, USA \\
}

\begin{abstract}
We present a multi-epoch hard X-ray analysis of \cyg\ in its soft state based on
four observations with \nustar. Despite the basic similarity of the observed spectra,
there is clear spectral variability between epochs. To investigate this variability, we
construct a model incorporating both the standard disk--corona continuum and
relativistic reflection from the accretion disk, based on prior work on \cyg, and apply
this model to each epoch independently. We find excellent consistency for the black
hole spin, and the iron abundance of the accretion disk, which are expected to
remain constant on observational timescales. In particular, we confirm that \cyg\
hosts a rapidly rotating black hole, $0.93 \lesssim a^{*} \lesssim 0.96$, in broad
agreement with the majority of prior studies of the relativistic disk reflection and
constraints on the spin obtained through studies of the thermal accretion disk
continuum. Our work also confirms the apparent misalignment between the inner
disk and the orbital plane of the binary system reported previously, finding the
magnitude of this warp to be $\sim$10--15\deg. This level of misalignment does not
significantly change (and may even improve) the agreement between our reflection
results and the thermal continuum results regarding the black hole spin. The
spectral variability observed by \nustar\ is dominated by the primary continuum,
implying variability in the temperature of the scattering electron plasma. Finally, we
consistently observe absorption from ionized iron at $\sim$6.7\,keV, which varies in
strength as a function of orbital phase in a manner consistent with the absorbing
material being an ionized phase of the focused stellar wind from the supergiant
companion star.
\end{abstract}

\begin{keywords}
{Black hole physics -- X-rays: binaries -- X-rays: individual (\cyg)}
\end{keywords}

\section{Introduction}

\cyg, the first X-ray binary confirmed to host a black hole accretor (\citealt{Murdin71,
Tananbaum72, Gies86a}), is one of the best studied Galactic black hole binaries
(BHBs). This system consists of a $\sim$14.8\,\msun\ black hole (\citealt{Orosz11})
accreting from the stellar wind of a type O9.7Iab supergiant companion (HDE
226868; \citealt{Walborn73}), and is one of the closest black hole systems known
($D = 1.86$\,kpc; \citealt{Reid11}).

\begin{figure*}
\hspace*{-0.25cm}
%\epsscale{1.15}
%\plotone{./longterm2_orbit_6h.eps}
\centering
\includegraphics[width=17.5cm]{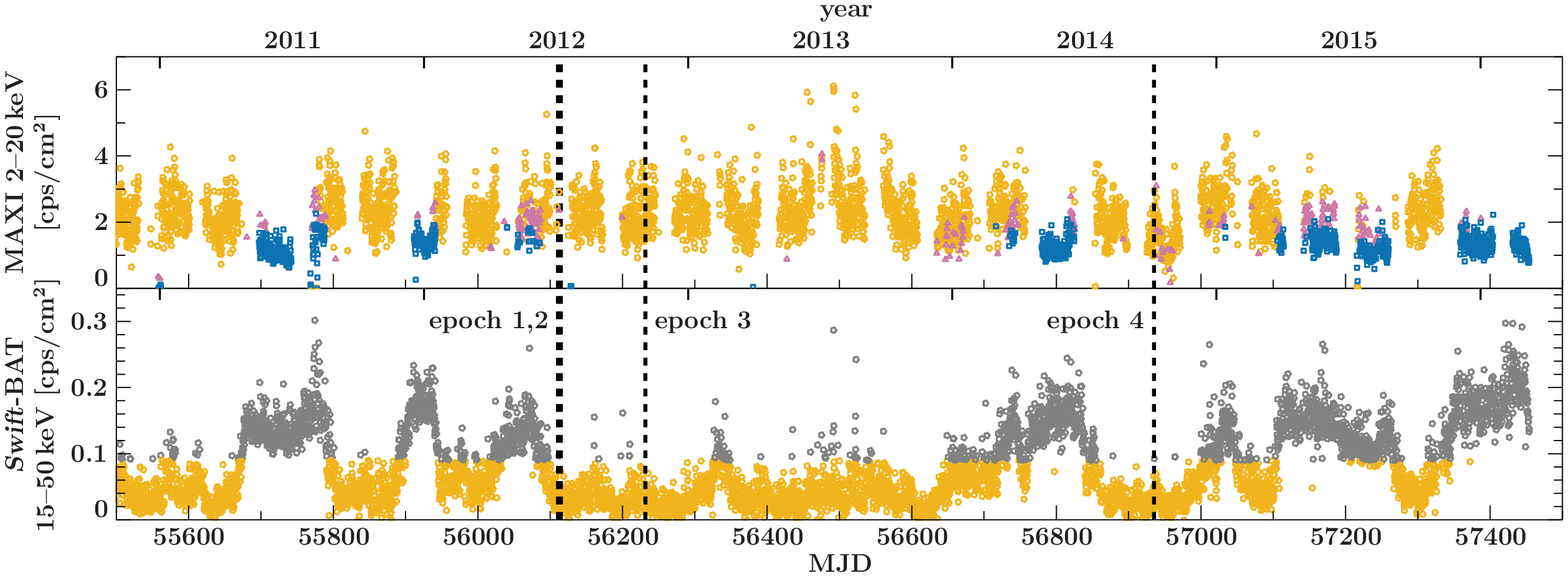}
\caption{
Long-term X-ray lightcurves (6\,hr bins) for \cyg\ observed with \maxi\ (\textit{top
panel}) and \swift\ BAT (\textit{bottom panel}) since $\sim$2011, with the \nustar\
observations considered in this work shown (dashed lnes). For the \maxi\ lightcurve,
soft states are shown in yellow, intermediate states in magenta, and hard states in
blue, following the definitions of \cite{Grinberg13cyg}. For the \swift\ BAT lightcurve,
soft states are again shown in yellow, and grey indicates either hard or intermediate
states (which can not easily be distinguished with the higher energy bandpass of
the BAT detectors).
}
\vspace{0.2cm}
\label{fig_longlc}
\end{figure*}

Besides its proximity, which makes it one of the brightest X-ray sources in the sky,
\cyg\ is an important system as it exhibits one of the best established examples of
relativistic reflection from the inner regions of its accretion disk (\eg
\citealt{Fabian89, Miller02, Reis10lhs, Duro11, Fabian12cyg}). This reflection is
produced when the accretion disk is irradiated by hard X-rays, and is dominated by
iron \ka\ emission ($\sim$6--7\,\kev, depending on ionization state) and a
characteristic high-energy continuum peaking at $\sim$20--30\,\kev\ (the `Compton
hump'; \eg \citealt{George91}). While the line emission is intrinsically narrow,
relativistic effects associated with the motion of the accreting material and the
extreme gravitational potential close to the black hole distort this emission into a
broad `diskline' profile (\citealt{Fabian89, kdblur, kerrconv, relconv}). Relativistically
broadened iron lines are observed in both Galactic BHBs and active galactic nuclei
(AGN), \eg \cite{Walton12xrbAGN}. Study of these distortions can provide geometric
constraints on both the inner accretion disk, and in turn the spin of the black hole
(\eg \citealt{Walton13spin, Reynolds14rev}), and the source of the illuminating hard
X-rays (\eg \citealt{Wilkins12}). The hard X-ray source is believed to be due to a
plasma of hot electrons, referred to as the `corona', which up-scatters the blackbody
emission from the accretion disk into a powerlaw-like high-energy continuum.

The \textit{Nuclear Spectroscopic Telescope Array} (\nustar; \citealt{NUSTAR})
covers the 3--79\,\kev\ band with unprecedented sensitivity above 10\,\kev. This
makes \nustar\ ideally suited for the study of relativistic disk reflection, as its broad
bandpass enables simultaneous measurements of both the iron emission and the
Compton hump. Furthermore, its triggered read-out means \nustar\ is also well
suited to deal with the high count-rates from Galactic BHBs, providing a clean, high
signal-to-noise measurement of the spectra of these sources. Since launch, \nustar\
has performed a series of observations to study relativistic reflection in this
population (\eg \citealt{Miller13grs, King14, Fuerst15, Tao15, Parker16}). In addition,
owing to its bandpass, \nustar\ is also well positioned to provide constraints on the
high-energy emission from the corona (\eg \citealt{Miller13grs, Natalucci14,
Fabian15}).

As part of its program to observe reflection in Galactic BHBs, \nustar\ has
performed a series of observations of \cyg\ (\citealt{Tomsick14, Parker15cyg}),
covering both the `soft' (disk-dominated) and `hard' (corona-dominated) accretion
states (see \citealt{Remillard06rev} for a review of Galactic BHB accretion states).
To date, these studies have found that \cyg\ hosts a rapidly rotating black hole. This
is in good agreement both with previous reflection-based results (\eg \citealt{Duro11,
Duro16, Fabian12cyg, Miller12}), as well as results from detailed study of the
blackbody disk emission (\eg \citealt{Gou11, Gou14}). The \nustar\ observations
also found evidence that the innermost accretion disk might be mis-aligned with the
orbital plane of the binary system (\citealt{Tomsick14, Parker15cyg}).

\begin{table}
  \caption{The soft state \nustar\ observations of Cygnus X-1 considered
  in this work.}
\begin{center}
\begin{tabular}{c c c c c}
\hline
\hline
\\[-0.175cm]
Epoch & Orbital & OBSID & Start Date & Exposure\tmark[b] \\
& Phase\tmark[a] & & & (ks) \\
\\[-0.2cm]
\hline
\hline
\\[-0.1cm]
1 & 0.26--0.37 & 00001011001 & 2012-07-02 & 14.4 (4.1) \\
\\[-0.225cm]
& & 00001011002 & & 5.2 (1.5) \\
\\[-0.2cm]
\hline
\\[-0.1cm]
2 & 0.97--0.09 & 10002003001 & 2012-07-06 & 12.5 (3.5) \\
\\[-0.2cm]
\hline
\\[-0.1cm]
3 & 0.85--0.99 & 30001011002 & 2012-10-31 & 11.0 (0.9) \\
\\[-0.225cm]
& & 30001011003 & & 5.7 (0.6) \\
\\[-0.225cm]
& & 10014001001 & & 4.6 (0.4) \\
\\[-0.2cm]
\hline
\\[-0.1cm]
4 & 0.46--0.59 & 30001011009 & 2014-10-04 & 22.6 (2.2) \\
\\[-0.2cm]
\hline
\hline
\\[-0.15cm]
\end{tabular}
\vspace{-0.2cm}
\flushleft
$^{a}$ Based on the ephemeris of \cite{Gies08}. \\
$^{b}$ The mode 6 contribution to the total exposure is given in parentheses.
\vspace*{0.3cm}
\label{tab_obs}
\end{center}
\end{table}

Soft state observations are particularly important for testing models of relativistic disk
reflection. It is widely accepted that in this state the accretion rate is high enough for
the disk to extend to the innermost stable circular orbit, which is necessary to
measure black hole spin. Here we present multi-epoch broadband X-ray observations
of \cyg\ with \nustar, with the purpose of examining the spectral variability and the
relativistic disk reflection exhibited during its soft, disk-dominated accretion state. The
paper is structured as follows: in section \ref{sec_red} we describe the \nustar\
observations and outline our data reduction procedure, in sections \ref{sec_spec} 
and \ref{sec_dis} we present our spectral analysis of these data and discuss the
results obtained, and finally in section \ref{sec_conc} we summarize our conclusions.

\begin{figure*}
\hspace*{-0.25cm}
\epsscale{0.56}
\plotone{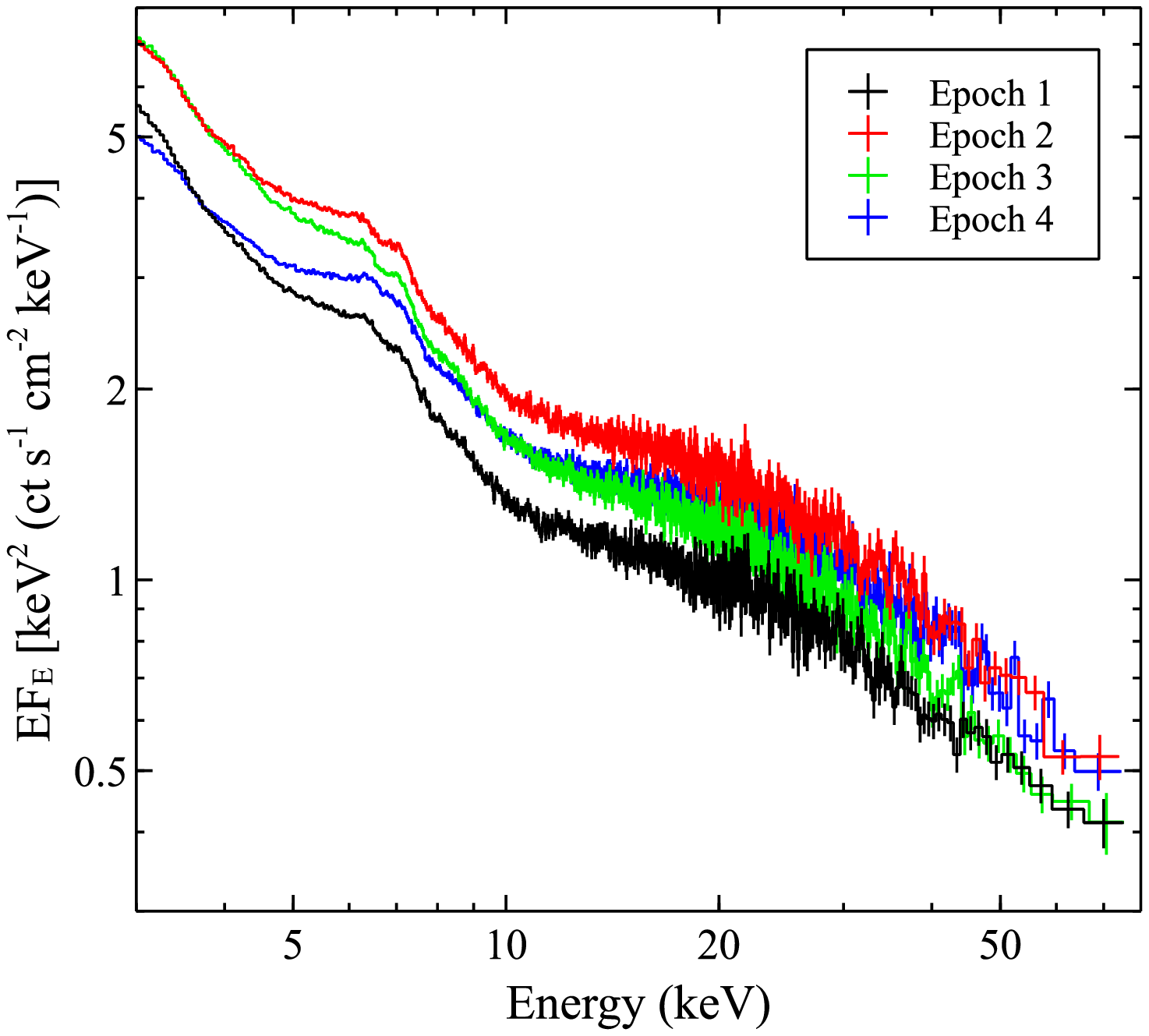}
\hspace*{0.5cm}
\plotone{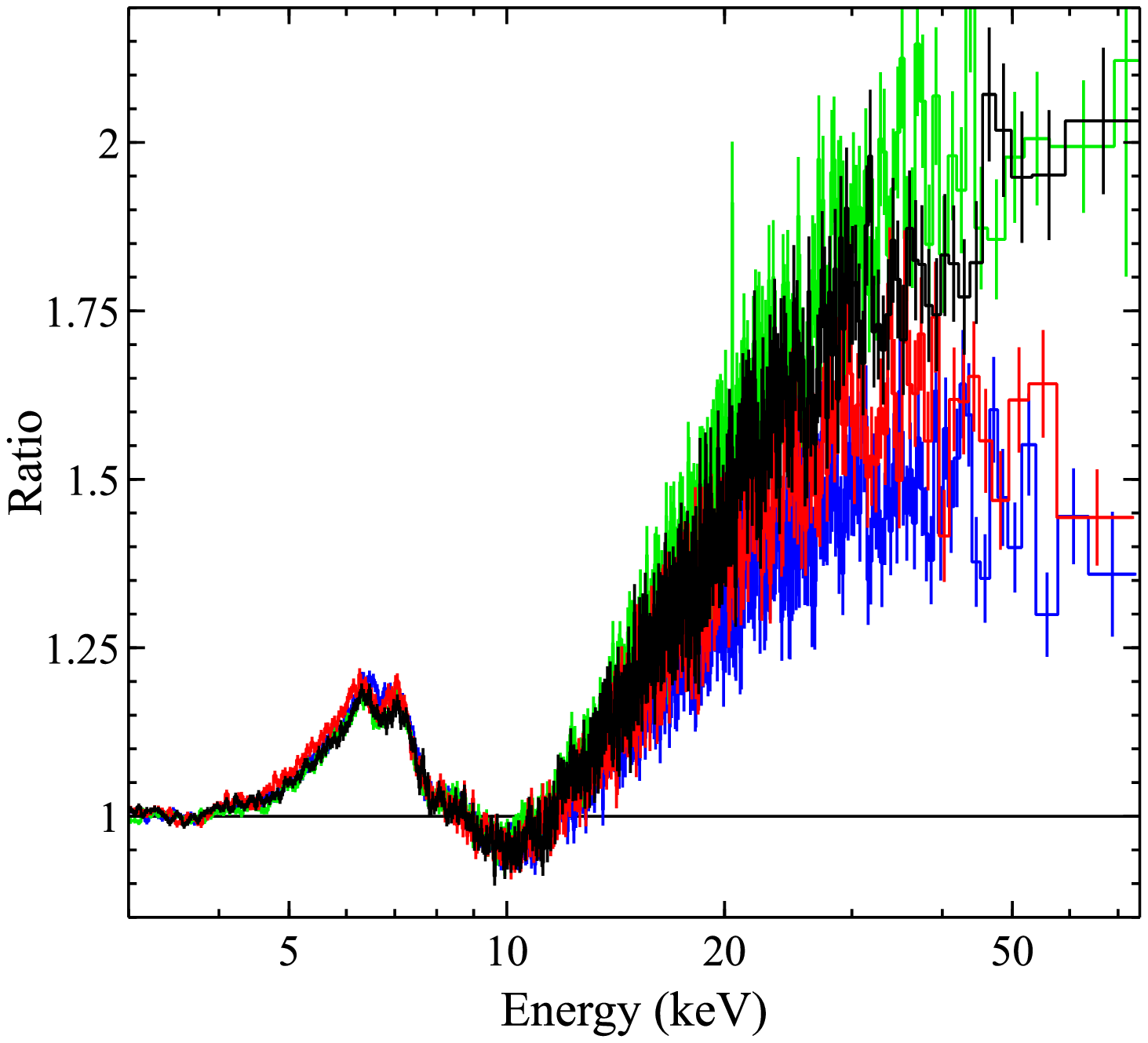}
\caption{
\textit{Left panel:} The X-ray spectra from the four \nustar\ observations of \cyg\ in its
soft state. The spectra have been unfolded through a model that is constant with
energy. \textit{Right panel:} Data/model ratios for these data after being fit with a
simple model consisting of an accretion disk and a high-energy powerlaw tail, fit to
the 3--4, 8--10 and 50--79\,keV bands. In each case, the residuals imply the
presence of a strong reflection component from the accretion disk (see
\citealt{Tomsick14}). Notably, a relativistically broadened iron line is seen in all four
\nustar\ soft state observations. While the residuals are broadly similar for all
epochs, there are visible differences at the highest energies probed by \nustar. For
both panels, the data have been further rebinned for visual purposes, and only the
FPMA data are shown for clarity.
}
\vspace{0.2cm}
\label{fig_soft_eeuf}
\end{figure*}

\section{Observations and Data Reduction}
\label{sec_red}

\nustar\ has performed soft state observations of \cyg\ at four epochs prior to 2015,
either as a science target (OBSIDs beginning with 3) or as an early mission
calibration target (OBSIDs beginning with 0 or 1); see Table \ref{tab_obs} for details.
Figure \ref{fig_longlc} shows how these observations relate to the long-term
behaviour seen from \cyg\ by both \maxi\ (\citealt{MAXI}) and \swift\ BAT
(\citealt{SWIFT_BAT}). Although multiple OBSIDs are listed for some epochs, these
are actually continuous observations. The science exposures from epoch 3 (OBSIDs
30001011002 and 30001011003) have already been presented in \cite{Tomsick14}.
However, immediately after those observations \nustar\ performed a further
calibration exposure. In this work, we utilize the data from all three of these
exposures. The rest of the observations included in this work are published here for
the first time.

We reduced the \nustar\ observations using the standard pipeline, part of the
\nustar\ Data Analysis Software v1.4.1 (\nustardas; part of the \heasoft\
distribution), adopting instrumental responses from \nustar\ CALDB version
20150316 throughout. The unfiltered event files were cleaned with the standard
depth correction, which significantly reduces the instrumental background at high
energies. Periods of Earth-occultation and passages through the South
Atlantic Anomaly were excluded. Source products were obtained from large
circular regions (radius $\sim$150$''$) for each of the two focal plane modules
(FPMs A and B). Owing to its brightness, there were no regions of the detector
on which \cyg\ was located that were free of source counts, so the
background was estimated from a blank region on the opposite detector (each
FPM carries four detectors in a square formation, see \citealt{NUSTAR}),
sufficiently far away from the position of \cyg\ to avoid any contribution from the
source. Although there are known to be variations in the background between
the detectors for each FPM, and so estimating the background from a different
detector to the source may formally introduce some systematic uncertainty,
these differences are typically only at the $\sim$10\% level (in the background
rate) at the highest energies of the \nustar\ bandpass (where the internal
detector background dominates; \citealt{NUSKYBKG}). All soft state spectra
obtained from \cyg\ are a factor of $\sim$3 or more above the background even
at these energies, and this increases very quickly towards lower energies, with
the source typically a factor of $\sim$1000 above the background at the low end
of the \nustar\ bandpass, so systematic effects due to background should have a
negligible influence on our results.

In addition to the standard ``science'' data (hereafter mode~1), we also reduce
the available ``spacecraft science'' data (referred to as mode 6) in order to utilize
the maximum possible exposure from each observation. In brief, this is the data
recorded during the periods of the observation for which the X-ray source is still
visible, but the star tracker on the optics bench cannot return a good aspect
solution, so the star trackers on board the spacecraft bus are utilized instead
(see Appendix \ref{app_mode6}). This provides an additional 10--40\% livetime,
depending on the observation (see Table \ref{tab_obs}). For epochs with multiple
OBSIDs, the data from each observation were reduced separately, and the
individual spectra from each FPM were then combined into a single average
spectrum from that epoch using \addascaspec, but we do not combine the data
from FPMA and FPMB, as recommended (\citealt{NUSTARcal}). The resulting
spectra from each FPM and each epoch were grouped to have a minimum of
50 counts per bin to facilitate the use of \chisq\ statistics in our analysis.

\section{Spectral Analysis}
\label{sec_spec}

In order to investigate the spectral variability exhibited by \cyg\ in its soft state, we
perform a detailed comparison of the \nustar\ data from each of the four epochs 
considered here. Throughout this work, our spectral analysis is performed with
\xspecv\ (\citealt{xspec}). All our models include a neutral Galactic absorption
component, modeled with \tbabs\ (\citealt{tbabs}). However, as \nustar\ is not
particularly sensitive to the typical column ($N_{\rm{H}} \sim 6 \times
10^{21}$\,cm$^{-2}$; \eg \citealt{Tomsick14}) owing to its bandpass (3--79\,\kev),
we fix the neutral absorption to $6 \times 10^{21}$\,cm$^{-2}$ for
simplicity.\footnote{Note that as outlined by \cite{Grinberg15cyg}, this is only
appropriate for the soft state, as the stellar wind is highly ionised and has little
effect on the neutral absorption (see also \citealt{Nowak11}).} We adopt the
abundances in \cite{tbabs} as our `solar' abundance set, as appropriate for
absorption by the Galactic interstellar medium, and as recommended we adopt
the cross-sections of \cite{Verner96}. Parameter uncertainties are quoted at
90\% confidence for one parameter of interest throughout this work, and we
account for residual cross-calibration flux uncertainties between the FPMA and
FPMB detectors by allowing multiplicative constants to float between them,
fixing FPMA to unity. The FPMB constants are always found to be within 5\% of
unity, as expected (\citealt{NUSTARcal}).

\begin{figure}
\hspace*{-0.35cm}
\epsscale{1.16}
\plotone{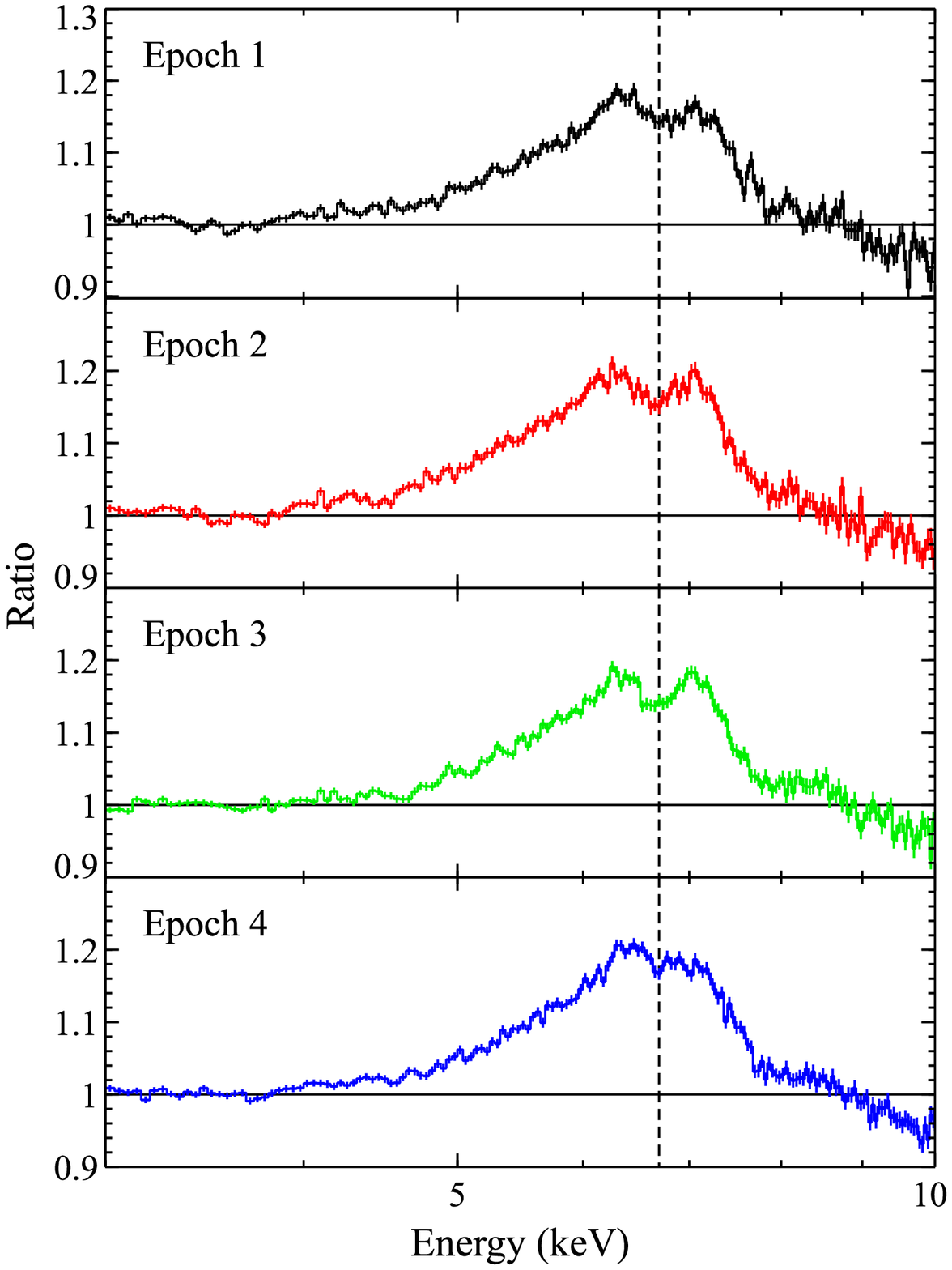}
\caption{
The same data/model residuals shown in Figure \ref{fig_soft_eeuf} (\textit{right
panel}), zoomed in on the iron \ka\ bandpass to highlight the iron emission. The line
profiles from each of the \nustar\ observation are all very similar. In addition to the
relativistically broadened iron emission, absorption of variable strength from ionized
iron can be seen at $\sim$6.7\,keV (the vertical dashed line shows $E=6.7$\,keV).
The data have been further rebinned for visual clarity.
}
\vspace{0.2cm}
\label{fig_feK_ratio}
\end{figure}

The \nustar\ spectra obtained from each of the four epochs are shown in the
left panel of Figure \ref{fig_soft_eeuf}. Although all four are broadly quite similar,
there are clear differences observed from epoch to epoch. This is further
demonstrated in the right panel of the same Figure, which shows data/model
ratios to a continuum model consisting of a multi-temperature blackbody
accretion disk (\diskbb; \citealt{diskbb}) with a high-energy powerlaw tail, fit to
the spectrum over the 3--4, 8--10 and 50--79\,\kev\ energy ranges to minimize
the influence on the model fit of the disk reflection spectrum known to be
present (\eg \citealt{Duro11, Fabian12cyg, Tomsick14}). There are clear
differences between epochs in the residuals to such a model at the highest
energies probed by \nustar. In contrast, the profile of the iron emission inferred
with this continuum appears to be rather stable; we show a zoom in on the iron
residuals in Figure \ref{fig_feK_ratio}. The broad iron emission is strong in the
soft state; adding a \relline\ component (\citealt{relconv}) to the simple
continuum model described above and fitting the 3--10\,keV band, we find
equivelent widths of $\mathrm{EW} \sim 300-330$\,eV for all four epochs. An
absorption feature associated with ionized iron is also visually apparent in the
residuals at $\sim$6.7\,\kev\ for some epochs, but its strength appears to be
variable.

%\pagebreak
\subsection{Model Setup}
\label{sec_model}

We construct a spectral model for \cyg\ building on the soft state analysis
presented in \cite{Tomsick14}. Similar to the majority of models considered in
that work, our model consists of an accretion disk component (\diskbb), a
high-energy powerlaw tail with an exponential high-energy cutoff representing
the coronal emission (\cutoffpl), a disk reflection component to account for the
iron emission and the reflected continuum observed at higher energies (using
the \xillver\ reflection code and the \relconv\ model to account for the relativistic
effects close to the black hole; \citealt{xillver} and \citealt{relconv} respectively),
and absorption from an ionized plasma to account for the 6.7\,\kev\ iron
absorption feature (\xstar; \citealt{xstar}).

Our choice of models is largely driven by pragmatic considerations. For the
reflected emission, we use the \xillver\ reflection model as this incorporates
the effects of viewing angle (i.e. the disk inclination, $i$) on the observed
reflection spectrum. We use the combination of \xillver\ and \relconv\ to
remain consistent with the approach of \cite{Tomsick14}. The \xillver\ family of
reflection models is calculated assuming a slab temperature of 10\,eV. This is
appropriate for the accretion disks around active galaxies, but much cooler
than the disk temperatures observed from X-ray binaries, and thus the
Compton broadening of the iron emission emerging from the accretion disk
will be underpredicted (\eg\ \citealt{refbhb}). Therefore, before application of
the \relconv\ model, we additionally smooth the \xillver\ model with a
Gaussian in order to approximate this additional broadening. This is similar
to Model 9 in \cite{Tomsick14}, but here rather than allow this broadening to
be an additional free parameter, we explicitly link it to the disk temperature,
\ie we assume that the free electrons within the disk also have this
temperature. The amount of broadening ($\sigma$) is then given by
$\sigma/E = \sqrt{2kT_{\rm{e}}/m_{\rm{e}}c^{2}}$, where $E$ is the line
energy, $k$ is Boltzmann's constant, $T_{\rm{e}}$ is the electron temperature,
$m_{\rm{e}}$ is the electron rest mass and $c$ is the speed of light
(\citealt{Pozdnyakov83}). Given the ionization of the disk inferred in
\cite{Tomsick14}, we also assume hydrogen-like iron is the dominant species
and adopt a typical line energy of $\sim$7\,\kev.

This choice of reflection model and our treatment of the additional Compton
broadening also drive our choice of models for the other continuum
components. For the disk emission, we use \diskbb\ as it explicitly has the
temperature as a free parameter ($kT_{\rm{in}}$), such that we can easily
determine the amount of Compton broadening to include. For the coronal
emission, we use a \cutoffpl\ model even though this is only an approximation
of the thermal Comptonization spectrum that may be expected from the
corona\footnote{There may also be a weak powerlaw contribution from
Comptonization by non-thermal electrons visible at extremely high ($\sim$MeV)
energies (\eg \citealt{McConnell02, Laurent11}), but this cannot be constrained
by our \nustar\ observations and so is not included in the model. When we refer
to the high-energy tail or the powerlaw component in this work, we are referring
to our approximation of the thermal Comptonization.}, as this is the input
spectrum assumed in the \xillver\ model, and we can therefore easily link the
parameters of the high-energy continuum that irradiates the disk (the photon
index $\Gamma$ and cutoff energy $E_{\rm{cut}}$) to those of the continuum
component included in the model. There has recently been some indication
from \nustar\ observations of Galactic BHBs that the disk might be irradiated by
a different hard X-ray continuum than that emitted along our line of sight
(\citealt{Parker15cyg, Fuerst15}), likely owing to a complex combination of
radial variations in the spectrum of the coronal emission, potential outflow
velocity gradients, and gravitational lightbending (\citealt{Fabian14}). However,
these have come from hard state observations, in which the corona might be
more significantly extended and/or outflowing (perhaps being associated with
the base of the jets launched in this state; \citealt{Markoff05, Miller12}), where
such variations might naturally be expected. In contrast, the corona is generally
expected to be more compact in the soft state (e.g. \citealt{Reis13lb}), and
radio jets are absent in this state (e.g. \citealt{Fender04}), thus the corona may
well be more static than in the hard state. We therefore assume that these
effects would be much less significant in the soft state, and we assume that the
continuum irradiating the disk is the same as that observed, noting that this has
previously worked well for the soft state (\citealt{Tomsick14}). Following
\cite{Garcia15}, we consider cutoff energies $E_{\rm{cut}} \leq 1$\,MeV.

\begin{table*}
  \caption{Results obtained for the free parameters in the disk reflection model
  constructed in this work.}
\begin{center}
\begin{tabular}{c c c c c c c}
\hline
\hline
\\[-0.175cm]
Model & \multicolumn{2}{c}{Parameter} & \multicolumn{4}{c}{Epoch} \\
Component & & & 1 & 2 & 3 & 4 \\
\\[-0.2cm]
\hline
\hline
\\[-0.1cm]
\diskbb\ & $kT_{\rm{in}}$ & [keV] & $0.405^{+0.010}_{-0.005}$ & $0.44 \pm 0.01$ & $0.47 \pm 0.01$ & $0.45 \pm 0.01$ \\
\\[-0.2cm]
& Norm & [$10^{4}$] & $15^{+2}_{-3}$ & $9^{+2}_{-3}$ & $6.2^{+0.6}_{-1.4}$ & $4.3^{+1.3}_{-0.2}$ \\
\\[-0.2cm]
\cutoffpl\ & $\Gamma$ & & $2.69 \pm 0.01$ & $2.58 \pm 0.02$ & $2.74^{+0.03}_{-0.04}$ & $2.56^{+0.02}_{-0.01}$ \\
\\[-0.2cm]
& $E_{\rm{cut}}$ & [keV] & $>600$ & $160^{+30}_{-30}$ & $280^{+110}_{-70}$ & $210^{+20}_{-40}$ \\
\\[-0.2cm]
& Norm (at 1\,keV) & [cts keV$^{-1}$ cm$^{-2}$ s$^{-1}$] & $5.2^{+0.3}_{-0.4}$ & $6.3^{+0.2}_{-0.3}$ & $8.0 \pm 0.2$ & $5.2 \pm 0.1$ \\
\\[-0.2cm]
\relconv\ & $a^{*}$ & & $0.948^{+0.006}_{-0.010}$ & $0.937^{+0.008}_{-0.007}$ & $0.939^{+0.008}_{-0.007}$ & $0.946 \pm 0.008$ \\
\\[-0.2cm]
& $i$ & [\deg] & $40.7^{+0.4}_{-0.8}$ & $38.2^{+0.9}_{-0.6}$ & $39.4^{+0.5}_{-0.7}$ & $40.8^{+0.5}_{-0.4}$ \\
\\[-0.2cm]
& \qin\ & & $>9.4$ & $>7.5$ & $>6.1$ & $>8.6$ \\
\\[-0.2cm]
& \rbr\ & [\rg] & $2.89^{+0.07}_{-0.04}$ & $2.91^{+0.42}_{-0.07}$ & $3.3^{+0.2}_{-0.4}$ & $2.63^{+0.09}_{-0.06}$ \\
\\[-0.2cm]
\xillver\ & $\log\xi_{\rm{refl}}$ & $\log$[\ergcmps] & $4.05^{+0.06}_{-0.02}$ & $4.02^{+0.03}_{-0.02}$ & $4.02^{+0.04}_{-0.02}$ & $3.82^{+0.03}_{-0.05}$ \\
\\[-0.2cm]
& $A_{\rm{Fe}}$ & [solar] & $4.2^{+0.3}_{-0.4}$ & $4.2^{+0.6}_{-0.4}$ & $4.3^{+0.4}_{-0.3}$ & $4.0 \pm 0.3$ \\
\\[-0.2cm]
& Norm & [$10^{-6}$] & $3.5^{+0.4}_{-0.8}$ & $4.0^{+0.4}_{-0.6}$ & $5.7^{+0.5}_{-0.7}$ & $4.0^{+1.0}_{-0.3}$ \\
\\[-0.2cm]
\xstar\ & \nh\ & [$10^{21}$ cm$^{-2}$] & $5.7^{+0.8}_{-0.6}$ & $9^{+2}_{-1}$ & $9.3^{+1.5}_{-0.9}$ & $4.5^{+0.7}_{-2.3}$ \\
\\[-0.2cm]
& $\log\xi_{\rm{abs}}$ & $\log$[\ergcmps] & $>3.9$ & $3.6^{+0.2}_{-0.1}$ & $3.7 \pm 0.2$ & $3.6^{+0.2}_{-0.1}$ \\
\\[-0.2cm]
& $v_{\rm{out}}$ & [\kmps] & $700^{+1500}_{-600}$ & $<600$ & $<800$ & $3500^{+300}_{-2000}$ \\
\\[-0.2cm]
\gauss\ & Line Flux\tmark[a] & [$10^{-12}$ \ergpcmsqps] & $5.6^{+1.5}_{-1.4}$ & $3.2^{+3.4}_{-2.8}$ & $3.4^{+2.7}_{-1.5}$ & $8.5^{+1.9}_{-1.2}$ \\
\\[-0.2cm]
& Equivalent Width & eV & $<40$ & $<100$ & $<14$ & $10^{+5}_{-4}$ \\
\\[-0.2cm]
\hline
\\[-0.1cm]
\chisq/DoF & & & 2049/1758 & 1978/1718 & 2089/1783 & 2080/1958 \\
\\[-0.2cm]
\hline
\hline
\\[-0.15cm]
\end{tabular}
\vspace{-0.2cm}
\label{tab_param}
\end{center}
$^{a}$ Calculated under the assumption that the narrow core of the line is present in all observations.
\vspace*{0.5cm}
\end{table*}

The other key free parameters for the reflection model are the black hole
spin ($a^*$), the iron abundance ($A_{\rm{Fe}}$) of the disk, its inclination
($i$) and its ionization parameter ($\xi = L/nR^{2}$, where $L$ is the ionizing
luminosity, $n$ the density and $R$ the distance to the ionizing source) and
lastly the radial emissivity profile of the reflected emission. We assume that
the disk extends to the innermost stable circular orbit (ISCO), as is also
generally expected for the soft state, and set the outer radius to the maximum
value allowed by the \relconv\ model (400\,\rg). The emissivity profile is
assumed to be a broken powerlaw, \ie $\epsilon(r) \propto r^{-q}$, where the
emissivity indices $q_{\rm{in, out}}$ differ on either side of some break radius
\rbr. Following \cite{Fabian12cyg}, we assume \qout\ to be 3 (as expected in
the simple Newtonian case; \citealt{Reynolds97feK}), and leave \qin\ and \rbr\
to be free parameters.

We also allow for the presence of a narrow ($\sigma=10$\,eV) Gaussian
emission line from neutral iron (6.4 keV) in the model, to account for X-ray
illumination of the wind launched from the massive stellar companion of \cyg.
Narrow iron emission lines are typically observed from high-mass X-ray
binaries (\eg \citealt{Torrejon10}), and the iron emission from \cyg\ has
previously been observed to be a composite of a broad and narrow
component (\eg \citealt{Miller02, Reis10lhs}). As such, in \xspec\ notation,
the final form of our model is: \tbabs\ $\times$ \xstar\ $\times ~($\diskbb\
$+$ \cutoffpl\ $+$ \gauss\ $+$ \relconv\ $\otimes$ \xillver$)$.

Finally, the \xstar\ model used for the ionized iron absorption is the same as
that used in \cite{Tomsick14}, for consistency. This is computed with \xstar\
version 2.2.1bg, with the ionization parameter and the column density of the
absorbing medium as free parameters. Elemental abundances are assumed
to be solar, and the model is computed with a density of $10^{12}$
cm$^{-3}$ and a turbulent velocity of 300\,\kmps\ (\citealt{Miller05, Hanke09}).
The input ionizing spectrum is based on a simple model for the soft state of
\cyg\ from \cite{Tomsick14}. In addition to the ionization and the column
density, we also allow the line-of-sight velocity of the absorbing medium
($v$) to be another free parameter in our analysis.

We apply this same model to the spectra from each of the four epochs
independently, in order to investigate the origin of the observed spectral
variability. During the course of our analysis, we found that the ionization
of the absorber was difficult to constrain, with statistically similar fits
($\Delta\chi^{2} \lesssim 10$) obtained for each observation for solutions
with $\log\xi_{\rm{abs}} \sim 3.8$ and $\log\xi_{\rm{abs}} \sim 5$ (where
$\xi$ is in units of \,\ergcmps); three of the four epochs marginally
preferred the lower ionization solution, while the remaining epoch (epoch
1) marginally preferred the higher ionization solution. We attribute the
multiple solutions found here to parameter degeneracies in the absorption
model that can occur when fitting a single, ionized iron absorption line (\eg
\citealt{King14}). However, given the level of variability apparent in Figure
\ref{fig_soft_eeuf}, the ionization is unlikely to differ by over an order of
magnitude between epochs. Furthermore, on investigation, we found that
the higher-ionization solution required the absorption to be significantly
redshifted, as it was fitting the feature at $\sim$6.7\,\kev\ with a blend of
\fexxv\ (6.67\,\kev) and \fexxvi\ (6.97\,\kev), which we deemed to be
unphysical. In our final analysis we therefore limit $\log\xi_{\rm{abs}} <
4$ and only allow the absorption to be blue-shifted (i.e. outflowing) to
exclude these higher-ionization solutions.

\subsection{Results}

\begin{figure}
\hspace*{-0.35cm}
\epsscale{1.16}
\plotone{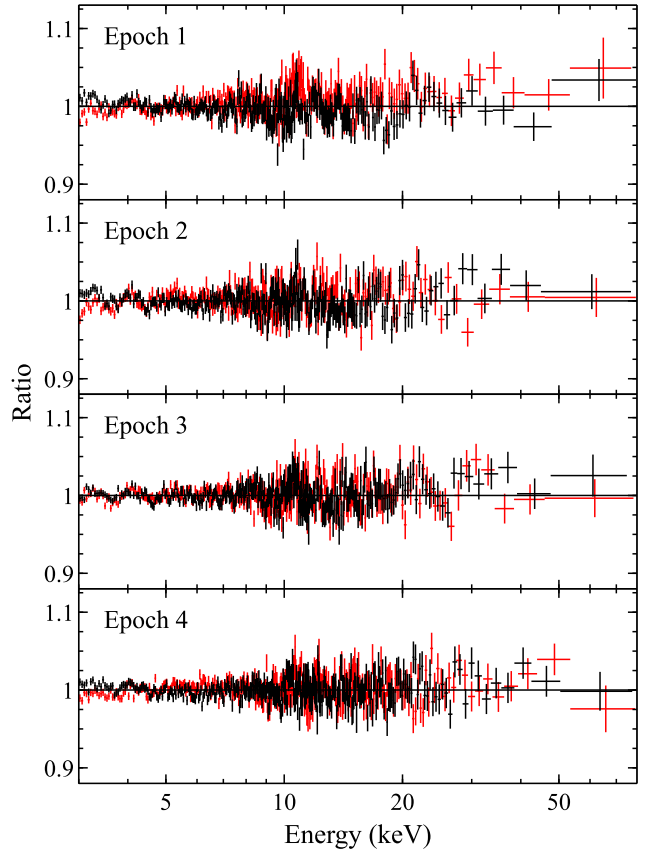}
\caption{
Data/model residuals for our final relativistic disk reflection model (see section
\ref{sec_model}). For each of the four epochs, FPMA data are shown in black, and
FPMB in red. Our model fits each of the soft state \nustar\ observations well.
As before, the data have been further rebinned for visual clarity.
}
\vspace{0.2cm}
\label{fig_mod_ratio}
\end{figure}

Our final model provides a good fit to all four soft state epochs observed by
\nustar\ to date. Data/model residuals are shown in Figure \ref{fig_mod_ratio},
and the best-fit parameters obtained are given in Table \ref{tab_param}. No
strong residuals remain across the entire \nustar\ bandpass.

\begin{figure}
\hspace*{-0.35cm}
\epsscale{1.16}
\plotone{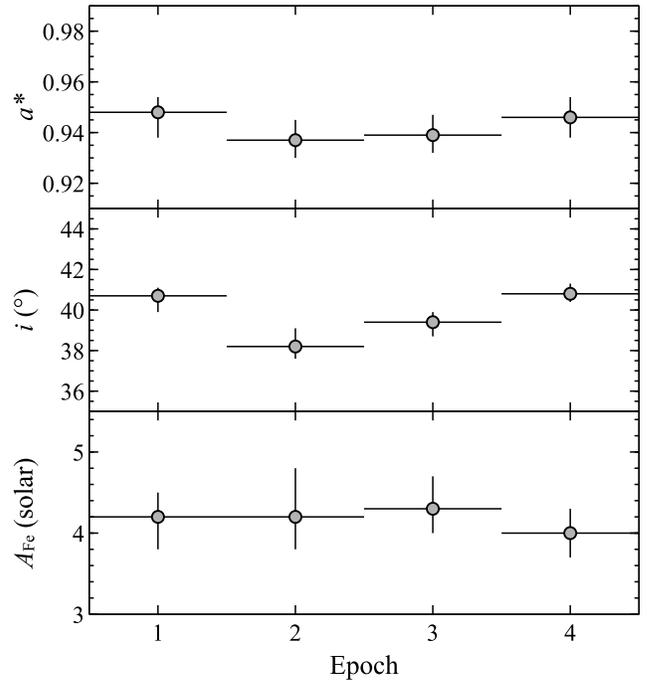}
\caption{
The results obtained for the black hole spin (\textit{top panel}), the disk inclination
(\textit{middle panel}) and the iron abundance (\textit{bottom panel}) from our
independent analysis of each epoch. These quantities should not change on
observable timescales, and indeed good consistency is observed between the
different epochs.
}
\vspace{0.2cm}
\label{fig_refl_param}
\end{figure}

Some noteworthy results are immediately apparent from a comparison of the
best-fit parameters obtained. First, the key disk reflection parameters that should
remain constant with time are indeed consistent across all four epochs (see
Figure \ref{fig_refl_param}). The consistency of the black hole spin confirms that
the inner disk radius remains constant during the soft state, as expected if the
disk extends all the way into the ISCO, and further reinforces the conclusion that
\cyg\ hosts a rapidly rotating black hole, as found by previous disk reflection
spectroscopy (\eg \citealt{Fabian12cyg}), and from study of the thermal continuum
(\eg \citealt{Gou11, Gou14}). In our model setup, the disk temperatures obtained
imply an expected Compton broadening of $\sigma \sim 0.3$\,\kev\ for all epochs,
similar to that obtained by \cite{Tomsick14} who allowed this to be an additional
free parameter (see their Model 9). This is clearly insufficient to explain the full
breadth of the observed iron emission, and does not change the requirement for
additional relativistic broadening.

Excellent consistency is also obtained for the iron abundance, which is found to
be super-solar. The abundance obtained here is higher than presented in
\cite{Tomsick14}, although we note that \cite{Tomsick14} make use of the
\reflionx\ model (\citealt{reflion}), while we use the \xillver\ reflection model. The
abundance obtained here is instead quite similar to that obtained by
\cite{Parker15cyg}, who also use the \xillver\ model in their analysis of a recent
broadband observation of \cyg\ by \suzaku\ and \nustar\ in the hard state
performed $\sim$5 months prior to our epoch 4.

\begin{figure}
\hspace*{-0.35cm}
\epsscale{1.16}
\plotone{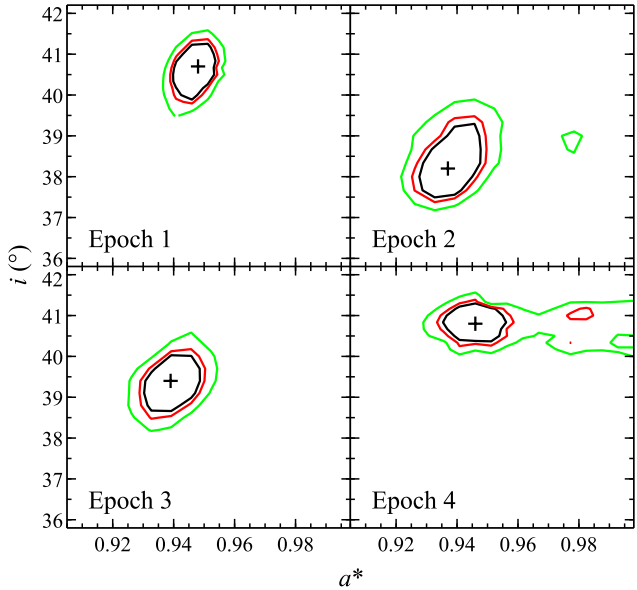}
\caption{
Two dimensional $\Delta$\chisq\ confidence contours for the black hole spin ($a^*$)
and the accretion disk inclination ($i$) for each epoch. The 90, 95 and 99\% confidence
intervals for two parameters of interest are shown in black, red and green, respectively.
}
\vspace{0.2cm}
\label{fig_spininc}
\end{figure}

The $\sim$40\deg\ disk inclination inferred with our model is also consistent across
all four epochs. As discussed in \cite{Tomsick14}, this is significantly larger than
the inclination at which we view the orbital plane of the system ($i_{\rm{orb}} =
27.1 \pm 0.8$\deg; \citealt{Orosz11}). We find that this apparent misalignment is
not confined to the single soft state epoch considered in \cite{Tomsick14}, and
does not show any evidence of varying on the timescales covered by our
observations. Furthermore, the inclination obtained here is similar to that found by
\cite{Parker15cyg}, so it does not appear to be confined to the soft state either. 
Prior studies of relativistic reflection from accreting black holes have found that the
black hole spin and the inclination inferred for the accretion disk often show some
degree of degeneracy, so in Figure \ref{fig_spininc} we compute two-dimensional
confidence contours for these parameters. While the degree of degeneracy seen
here varies somewhat between epochs, it is very mild, and not sufficient to reconcile
the measured inner disk inclination with the orbital plane measurement. Indeed, an
inner disk inclination of $\sim$27\deg\ is strongly disfavoured by our current model;
for illustration, we fit epoch 1 with the inclination fixed to 27.1\deg, resulting in the fit
degrading by $\Delta\chi^{2}$ = 450 and noticeable residuals around the blue wing
if the iron line. However, it is worth noting that even with this fit, the spin inferred for
\cyg\ is high ($a^* > 0.91$).

\begin{figure}
\hspace*{-0.35cm}
\epsscale{1.16}
\plotone{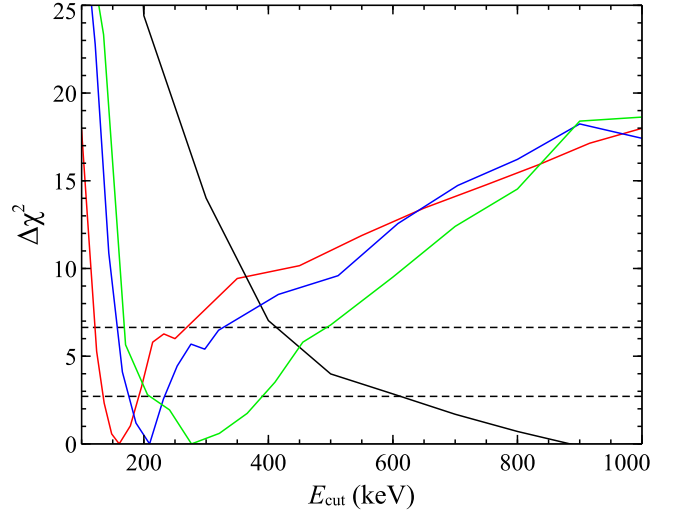}
\caption{
The $\Delta$\chisq\ confidence contours for the high-energy cutoff ($E_{\rm{cut}}$)
for each of the four soft state \nustar\ observations. The contours for epochs 1, 2, 3
and 4 are shown in black, red, green and blue, respectively. The horizontal dashed
lines represent the 90 and 99\% confidence levels for a single parameter of interest.
}
\vspace{0.2cm}
\label{fig_ecut}
\end{figure}

In contrast to the stability of the disk reflection parameters, we find evidence for
variability in the high-energy continuum parameters. Most notably the cutoff
energy -- which is indicative of the electron temperature assuming the high-energy
continuum is dominated by thermal Comptonization -- is found to vary between
epochs by at least a factor of $\sim$3--4 (ranging from $\sim$150 up to
$>$600\,\kev; see Figure \ref{fig_ecut}), resulting in the differences at the highest
energies probed by \nustar\ seen in Figure \ref{fig_soft_eeuf}. The range of values
obtained here is broadly similar to recent results obtained for the soft state with
the high-energy detectors aboard \integral, although these were averaged over
several epochs (\citealt{Jourdain14, Rodriguez15}).

\begin{figure}
\hspace*{-0.35cm}
\epsscale{1.16}
\plotone{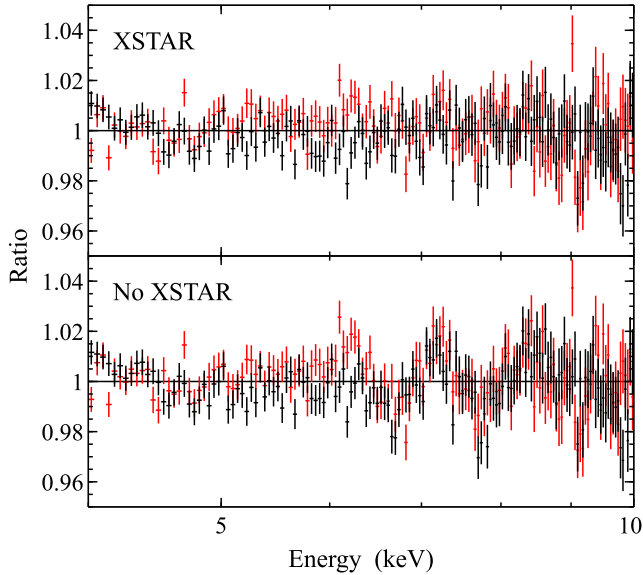}
\caption{
Data/model residuals for epoch 4, zoomed in on the iron \ka\ band. The top panel
shows our best-fit reflection model, including ionized iron absorption (modeled with
XSTAR), and the bottom panel shows the best-fit for the same model but with the
ionized absorption removed. Although the absorption is much weaker during this
epoch, it is still required to fit the data. As before, the data have been further
rebinned for plotting purposes; colors are as in Figure \ref{fig_mod_ratio}.
}
\vspace{0.2cm}
\label{fig_o4_abs}
\end{figure}

We also confirm the visual indication from Figure \ref{fig_feK_ratio} that the strength
of the absorption at $\sim$6.7\,\kev\ is variable, with the column density obtained for
the ionized absorbing medium varying from epoch to epoch. We also note that
despite not being visually apparent, this absorption is strongly required in the fourth
epoch. Excluding the \xstar\ component from the model for this epoch results in a
significantly worse fit ($\Delta\chi^{2} = 109$ for three fewer free parameters), and
as shown in Figure \ref{fig_o4_abs} leaves clear absorption residuals. It is only after
accounting for the relativistic disk reflection that the imprint of this absorption
becomes obvious.

Finally, with regard to the narrow core of the iron emission, we find that this is
strongly required for epoch 4 ($\Delta\chi^{2} = 78$ for one additional free
parameter), and also gives a moderate improvement for epoch 1 ($\Delta\chi^{2} =
14$). In contrast, for epochs 2 and 3, including the narrow core does not really
provide a meaningful improvement to the fit ($\Delta\chi^{2} = 4$ and 7,
respectively). Nevertheless, under the assumption that the narrow core is present
at each epoch, we still calculate the line flux, which we find to show some level of
variability between epochs (unsurprisingly being strongest during epoch 4). We also
compute the equivalent width using the {\small EQWIDTH} command in \xspec, but
only obtain upper limits for epochs 1, 2 and 3 owing to a combination of the weak
statistical improvement this feature provides and the complexity of the underlying
continuum model.

\section{Discussion}
\label{sec_dis}

We have presented a multi-epoch, hard X-ray (3--79\,keV) analysis of the canonical
black hole high-mass X-ray binary \cyg\ in its soft state with \nustar. \nustar\
has observed \cyg\ in this state on four separate epochs prior to 2015. One of the
great advantages \nustar\ has when studying sources as bright as \cyg\ is that it
does not suffer from pile-up, owing to its triggered read-out (\citealt{NUSTAR}), and
thus provides a clean, high signal-to-noise view of the X-ray spectrum. Although we
observe some variation between the different epochs (see Fig. \ref{fig_soft_eeuf}),
these observations all show broadly similar spectra with strong disk emission
(peaking below the \nustar\ bandpass; \citealt{Tomsick14}) and a steep high-energy
tail ($\Gamma \sim 2.6$, consistent with the soft state criterion outlined by
\citealt{Grinberg13cyg}). In addition to this continuum emission, each of the four
epochs reveals a clear contribution from relativistic reflection from the inner accretion
disk (see Figures \ref{fig_soft_eeuf}, \ref{fig_feK_ratio}), enabling us to probe the
inner accretion geometry, as well as ionized absorption from ionized iron (see
Figures \ref{fig_feK_ratio}, \ref{fig_o4_abs}).

\begin{figure}
\hspace*{-0.35cm}
\epsscale{1.16}
\plotone{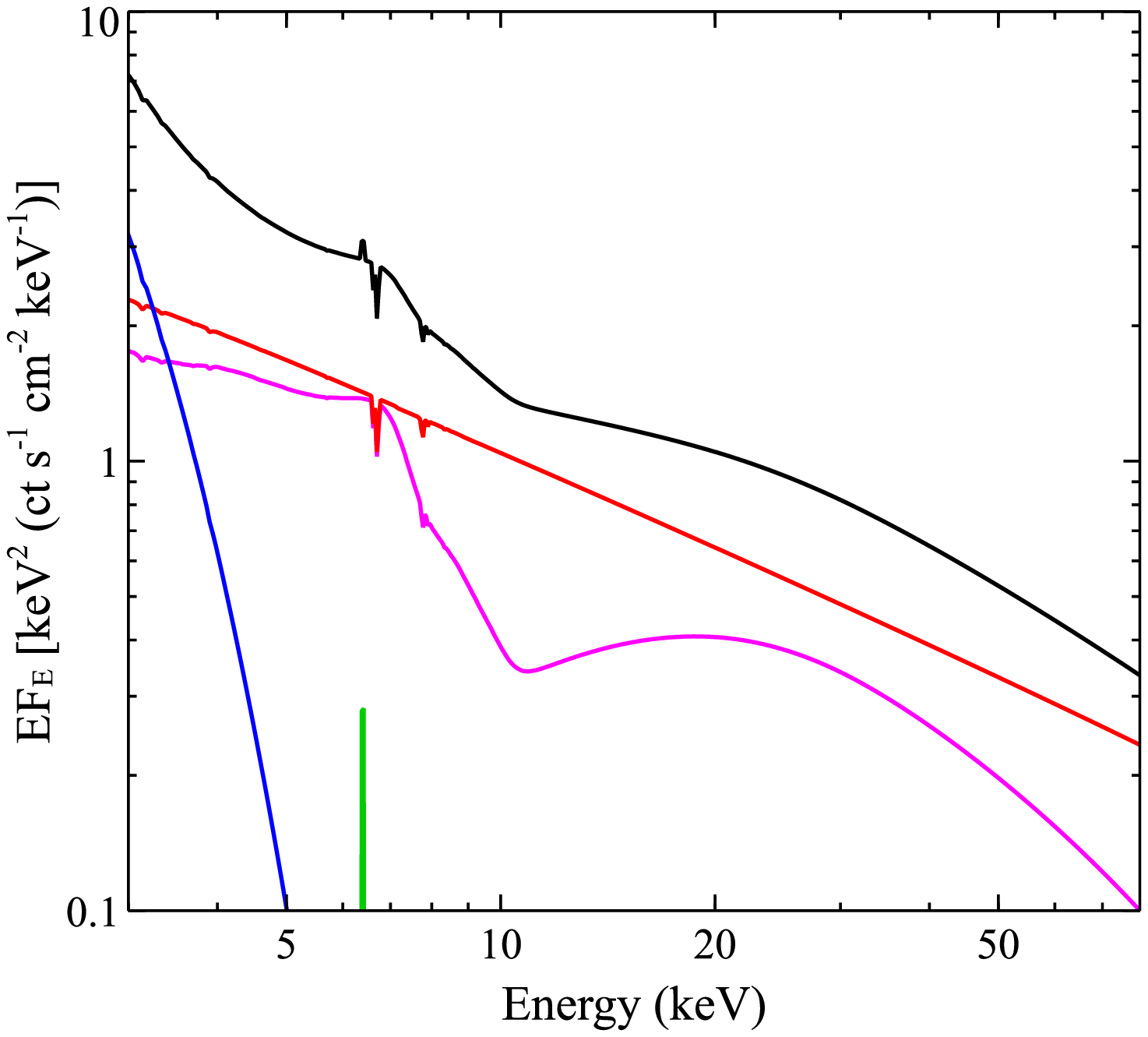}
\caption{
The best-fit disk reflection model obtained for our epoch 1. The total model is
shown in black, and the relative contributions across the \nustar\ bandpass from
the accretion disk (blue), the high-energy powerlaw tail (red), the disk reflection
(magenta) and the narrow core of the iron emission (green) are also shown.
}
\vspace{0.2cm}
\label{fig_o1_mod}
\end{figure}

Building on the original work of \cite{Tomsick14}, who present a detailed analysis of
the \nustar\ data that comprise the majority of our epoch 3, we construct a relativistic
disk reflection model and apply this to each of the four epochs independently in 
order to investigate the spectral variability observed between them. The best-fit
model obtained for epoch~1 is shown in Figure \ref{fig_o1_mod}, showing an
example of the relative contributions across the \nustar\ bandpass from the different
model components.

\subsection{Black Hole Spin and the Inner Accretion Disk}

Even though there is clear broadband spectral variability, the profile of the iron
emission inferred from simple continuum models is found to be very similar for all
epochs (Figure \ref{fig_feK_ratio}), implying that the geometry of the accretion disk
is stable throughout these observations. Indeed, the quantitative results obtained
from our disk reflection modeling show excellent consistency for the key parameters
that should not vary on observational timescales (black hole spin, disk inclination,
iron abundance; Figure \ref{fig_refl_param}). This is similar to the recent multi-epoch
analysis of the large \nustar+\xmm\ campaign on the active galaxy NGC\,1365
(\citealt{Walton14}), where excellent consistency was also found for the key
reflection parameters. It is worth noting that in between these soft state observations
\cyg\ underwent at least one transition to a hard state (\eg between epochs 3 and 4,
see Figure \ref{fig_longlc}; \citealt{Parker15cyg}), so we are not simply observing
one continuous, uninterrupted soft state, but rather a configuration that \cyg\ has
returned to on more than one occasion.

Our multi-epoch results present a consistent picture, that \cyg\ hosts a rapidly
rotating black hole. Our analysis constrains the spin of \cyg\ to be $0.93 \lesssim
a^{*} \lesssim 0.96$ (based on the 90\% statistical uncertainties on the
measurements from each epoch). The quantitative results obtained for the spin
here (Table \ref{tab_param}) are consistent with the majority of the constraints
obtained from previous reflection modeling, both from the \nustar\ observations
analysed prior to this work (\citealt{Tomsick14, Parker15cyg}), and from analyses
based on observations with \xmm, \rxte, \suzaku\ and \integral\ (\eg \citealt{Duro11,
Duro16, Fabian12cyg, Miller12}), despite these works using a range of different
reflection models fit over a variety of different bandpasses, and even considering
different accretion states. This is also in good agreement with the results obtained 
through detailed study of the thermal accretion disk emission, which also find a high
spin for \cyg\ (\citealt{Gou11, Gou14}). The consistency between the two techniques
leads us to conclude that these measurements are robust, and that \cyg\ hosts a
rapidly rotating black hole. While there are still only a few \nustar\ spin constraints
for Galactic BHBs, owing in large part to the relatively short mission lifetime at the
time of writing, we note that there is also excellent consistency between the \nustar\
reflection results and the thermal continuum for the only other case we are aware of
where both are available, GRS\,1915+105 (\citealt{McClintock06, Miller13grs}).

The inclination we obtain for the inner disk is $\sim$40\deg\ for all epochs, broadly
consistent with the previous \nustar\ results but significantly larger than the orbital
inclination reported by \cite{Orosz11}, $i_{\rm{orb}} = 27.1 \pm 0.8$\deg, confirming
the discrepancy discussed by both \cite{Tomsick14} and \cite{Parker15cyg}.
Currently the best interpretation for this discrepancy relates to a warp in the
accretion disk. This would need to be of order $\sim$10--15\deg\ based on our
results, which may be plausible given the binary population synthesis work of
\cite{Fragos10}. Furthermore, the 3D simulations of \cite{Nealon15} suggests such
warps/misalignments may remain stable, as our results suggest. If this warp is real,
in principle this would mean that adopting the orbital inclination when estimating the
black hole spin from the thermal accretion disk continuum would not be correct.
However, \cite{Gou11} show the inferred spin as a function of disk inclination (see
their Figure 5), and for the inclination found here the result inferred for the spin of
\cyg\ changes only marginally. In fact, adopting our inclination improves the formal
quantitative agreement between the spin inferred from the thermal continuum,
which would change from $a^{*} \sim 0.998$ to $a^{*}\sim 0.96$, and the constraints
obtained in this work (see above). However, such differences are likely small in
comparison to the systematic errors associated with such measurements (with both
techniques), so we consider the agreement to be excellent regardless of whether
the inferred warp is real or not.

Finally, we find that the accretion disk has a super-solar iron abundance. While this
is driven in part by the strength of the iron emission (see section \ref{sec_spec}), it
is also influenced by the Compton hump, as the iron absorption helps determine the
curvature of the spectrum on its red (low-energy) side. For illustration, we also fit the
data from epoch 2 with our model just above 10\,keV. The parameters of the \diskbb\
and \xstar\ components were fixed at their best-fit  values, as were the inclination and
ionization state of the accretion disk, since the data considered are not particularly
sensitive to these parameters. The constraint on the iron abundance is considerably
weaker, but we still find a lower limit of Fe/solar $>$ 4.4 from the data above 10\,keV
alone, consistent with the results for the full \nustar\ bandpass. Interestingly, these
data also provide a weak constraint on the spin of $a^{*} > 0.3$, also consistent with
the full band fits.

A high iron abundance is in qualitative agreement with other studies in which it was
allowed to vary, even if there is some quantitative tension. Here, the quantitative
results are reasonably well split by the reflection model utilized, and when the same
model is compared good agreement is seen. Results obtained using the \reflionx\
code (\citealt{reflion}) generally tend to return Fe/solar $\sim$ 2 (\eg\ \citealt{Duro11,
Fabian12cyg, Tomsick14})\footnote{One notable exception is the recent work by
\cite{Duro16}, who find an abundance of Fe/solar $\sim$3--4, despite using the
\reflionx\ model.}, while results obtained with \xillver-based models (\citealt{xillver,
relxill}) tend to return Fe/solar $\sim$ 4 (this work, \citealt{Parker15cyg}). Although
this may in part be due to the different solar abundance sets adopted by the two
models -- the \reflionx\ models adopt the abundances given in \cite{Morrison83}
while the \xillver\ models adopt those given in \cite{Grevesse98} -- in terms of their
Fe/H number ratios the \xillver\ model is only lower by $\sim$30\% (see Table 1 in
\citealt{xillver13}), insufficient to be the sole cause of this discrepancy. While we
conclude that current models do imply that the \cyg\ system has a super-solar iron
abundance (under the assumption that the system is chemically homogeneous),
the exact abundance still appears to be subject to substantial systematic
uncertainties. We note that \cite{Hanke09} also found evidence for super-solar
abundances for elements including iron in the \cyg\ system through high-resolution
studies of the absorption in the stellar wind along our line-of-sight. Dedicated
studies undertaking detailed consideration of both the reflection and the absorption
simultaneously, utilizing simultaneous broadband and high-resolution observations
(with \eg \hitomi; \citealt{ASTROH_tmp}) will be required to help further address this
issue in the future. We therefore defer a detailed discussion regarding the origin of
the super-solar Fe abundance inferred to such future work. However, given that
\cyg\ accretes from the stellar wind of its companion one interesting possibility is
that this over-abundance is related to the `first ionisation potential effect' seen in
some phases of the Solar corona/wind, in which elements with first ionisation
potential below $\sim$10\,eV (including Fe) show enhanced abundances (see
\citealt{Laming15rev} for a recent review). The exact cause of this effect remains
an active area of research. Alternatively, or in combination, metal enrichment from
the supernova that produced the black hole powering the \cyg\ system could also
result in an enhanced iron abundance.

% NOTE: Grevesse96 and Grevesse98 abundances sets are the same.

\begin{figure*}
%\hspace*{-0.35cm}
\hspace*{-0.5cm}
\epsscale{0.55}
\plotone{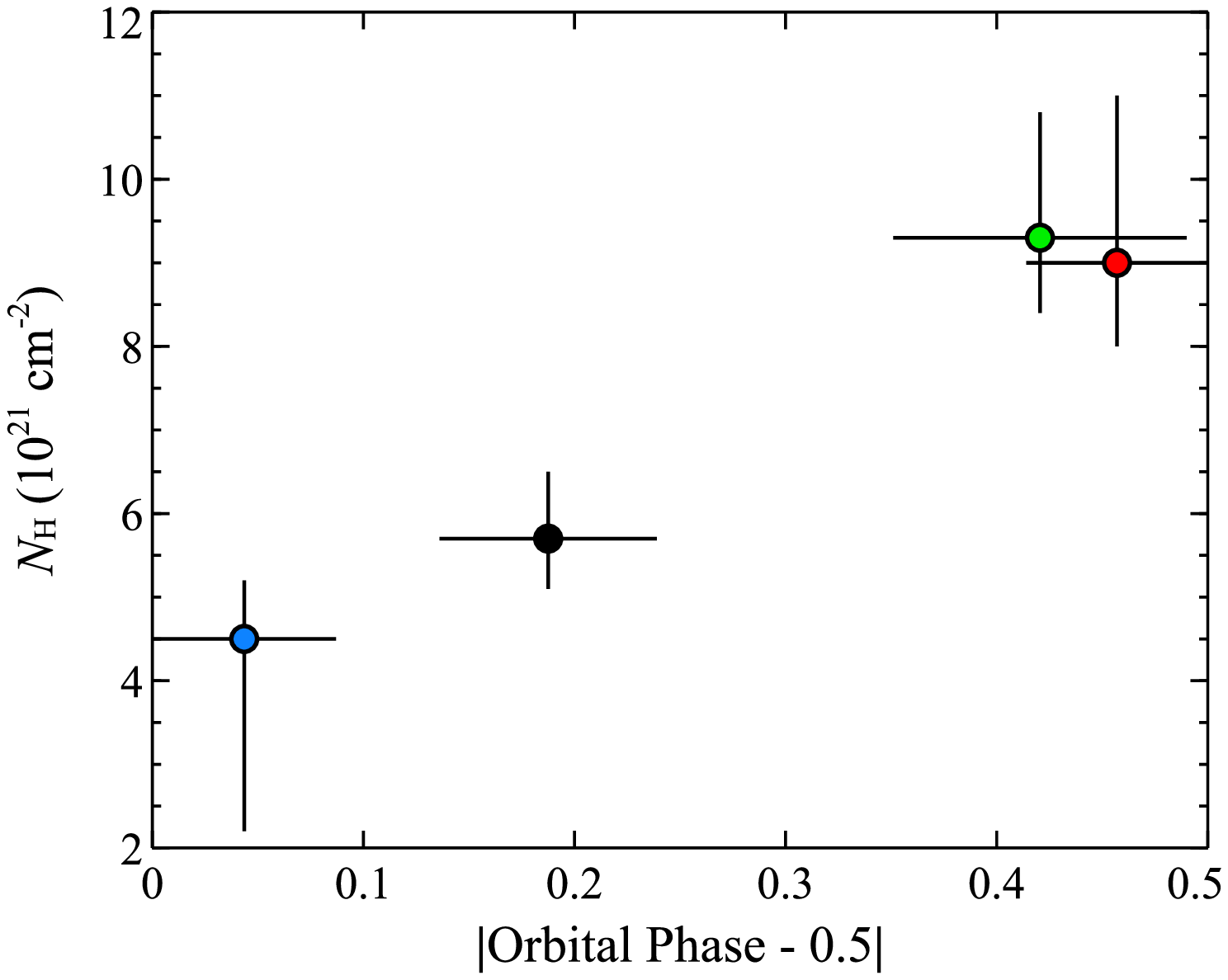}
\hspace*{0.525cm}
\plotone{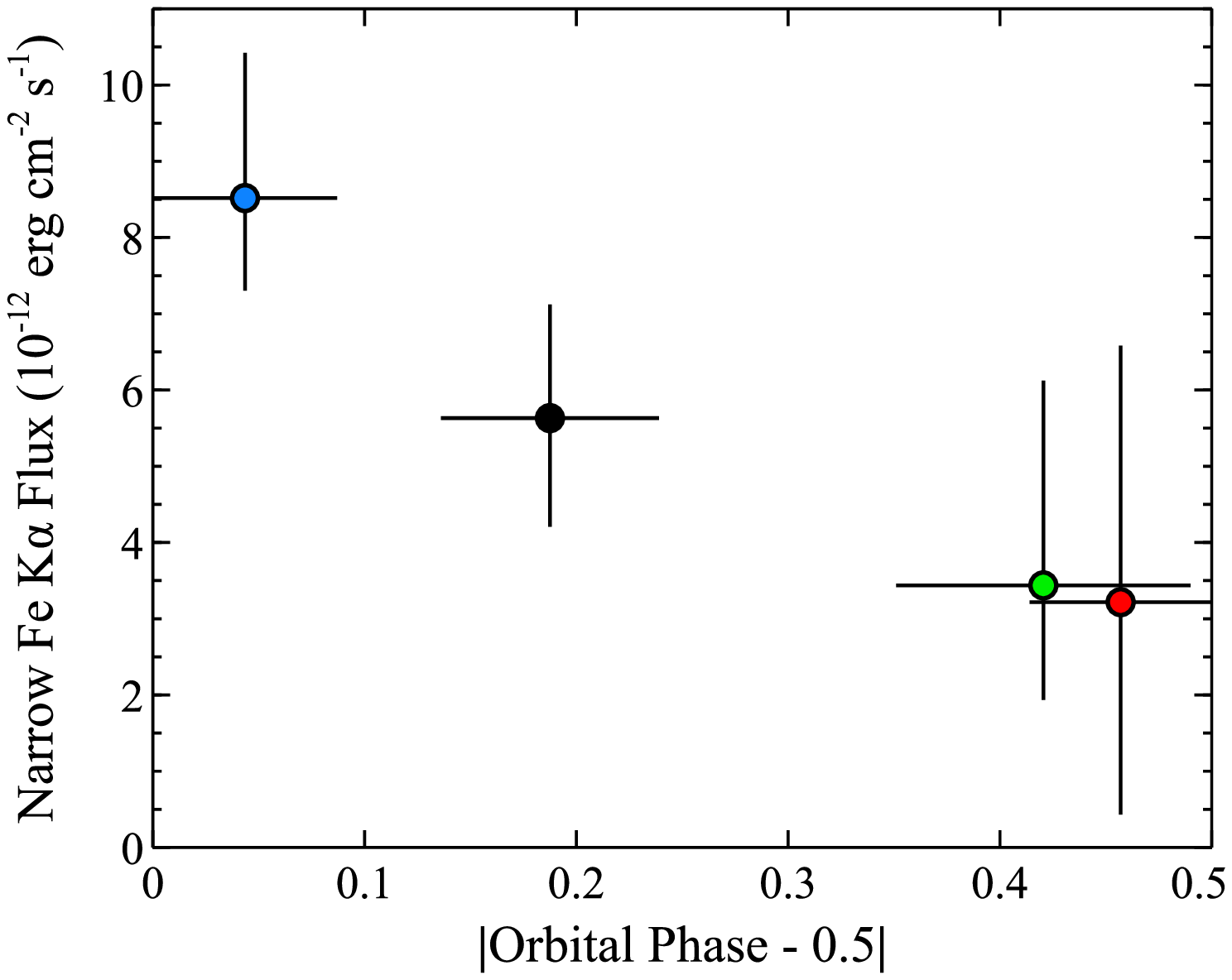}
\caption{
The strength of the ionized iron absorption (\textit{left panel}) and the flux
of the narrow core of the iron emission (\textit{right panel}; computed assuming this
component is present at each epoch) as a function of the orbital phase at which the
\nustar\ observation was performed. Here we plot the orbital phase in terms of the
separation from superior conjunction (i.e. the point at which the companion star is
on the far side of the black hole; $\phi_{\rm{orb}} = 0.5$). As the orbital phase
moves away from this point, the strength of the absorption increases. However, the
absorption is not completely absent at $\phi_{\rm{orb}} \sim 0.5$. We also find that
the narrow line emission is strongest at this point, although the evidence for this
variation is much more marginal. The data points are colored by epoch, following
the color scheme in Figures \ref{fig_soft_eeuf} and \ref{fig_feK_ratio}.
}
\vspace{0.2cm}
\label{fig_orbvar}
\end{figure*}

\subsection{The High-Energy Emission}

In contrast to the relative stability of the reflection, the primary continuum is variable,
driving the observed spectral variability. We formally see some evidence for changes
in the temperature of the accretion disk between epochs, but given that the disk
emission peaks outside the \nustar\ band we treat this with some caution. The
\nustar\ band primarily covers the high-energy tail, in which we also see variations,
both in terms of overall slope but in particular at the highest energies probed by
\nustar\ (see Figure \ref{fig_soft_eeuf}). Although we model this with a simple
phenomenological cutoff powerlaw model for practical reasons, this tail is widely
expected to originate through Compton up-scattering of accretion disk photons by a
corona of hot electrons (e.g. \citealt{Haardt91}). In this case, the energy of the
exponential cutoff of the powerlaw component acts as a proxy for the temperature
of the scattering electrons.\footnote{We stress however that the conversion between
$E_{\rm{cut}}$ and electron temperature is not necessarily linear, particularly when
the inferred cutoff energy is significantly outside the observed bandpass, as an
exponential cutoff is only a rough approximation of the high-energy cutoff produced
by thermal Comptonization, which curves faster with energy once it starts falling
away (\citealt{Zdziarski03}). While a higher energy cutoff does indicate a hotter
electron population, a factor of $\sim$4 change in $E_{\rm{cut}}$ does not
necessarily imply exactly the same level of variation in the electron temperature.
See \cite{Fabian15} and \cite{Fuerst16} for more discussion.}

We see significant variation in the cutoff energy $E_{\rm{cut}}$ (see Figure
\ref{fig_ecut}), which changes by a factor of $\gtrsim$3--4, driving the variability
seen at the top of the \nustar\ band. This indicates that the temperature of the
scattering electrons varies between epochs. Recently \cite{Fabian15} investigated
the implications of the high-energy cutoff measurements obtained to date for
accreting black holes by \nustar\ (for both AGN and BHBs), finding that the results
suggested that the X-ray coronae around these objects are likely in a regime in
which pair production/annihilation is important. Our results suggest that this is
persistently the case for the soft state of \cyg, and we speculate that the variability
observed in the high-energy cutoff (and thus inferred for the electron temperature)
here might indicate the source approaching and receding from the point of
catastrophic pair production discussed by \citet[and references therein]{Fabian15}.

\subsection{Narrow Absorption and Emission}

In addition to the high-energy continuum, we also see variability in the strength of
the ionized iron absorption at $\sim$6.7\,keV, with some evidence for an orbital
dependence. The absorption is stronger at orbital phases when the system is close
to inferior conjunction (i.e. the black hole is on the far side of the binary system from
our perspective, $\phi_{\rm{orb}} \sim 0.0$ and $\phi_{\rm{orb}} \sim 1.0$), and
weaker when the system is close to superior conjunection (i.e. $\phi_{\rm{orb}} \sim
0.5$; see Figure \ref{fig_orbvar}, left panel). This is consistent with the absorption
arising in an ionized phase of the stellar wind of the supergiant companion from
which \cyg\ accretes, as discussed by \eg \cite{Miller05, Hanke09, Nowak11}.
However, despite not being immediately visually obvious, the absorption is not
completely absent in our observation that is closest to superior conjunction (epoch
4), suggesting that the stellar wind still pollutes some of the region on the far side
of the binary system (with respect to the stellar companion). \cite{Grinberg15cyg}
studied the time evolution of the neutral absorption in detail, primarily during the
hard state, and also found similar results, specifically evidence for absorption by
the (clumpy) stellar wind even at $\phi_{\rm{orb}} \sim 0.5$. Interestingly, epoch 4
is also where we see the strongest (although still fairly marginal) evidence for a
blueshift in the absorption. It has long been suggested that some portion of the
stellar wind from HDE 226868 is focused towards the black hole (\eg
\citealt{Gies86b, Friend82, Miller05, Hanke09}), a scenario which likely provides a
natural explanation for all of these results (see Figure 3 in \citealt{Miller05} for a
suggested geometry).

A potential alternative origin for the absorption could be that this arises in a disk
wind, similar to those seen in the soft states of other Galactic BHBs (\eg\
\citealt{Miller06a}). However, we do not consider this to be particularly likely. The
orbital modulation indicated in Figure \ref{fig_orbvar} would not be expected in
this scenario (although with so few data points it is difficult to completely exclude a
coincidental alignment of orbital phase and the strength of any disk wind).
Furthermore, this absorption is also seen in the hard state (\citealt{Parker15cyg,
Miskovicova16}), which is not typical of other BHB systems that do exhibit
absorption from soft state disk winds (\eg\ \citealt{Neilsen09}). In particular, using
\chandra\ HETG observations, \cite{Miskovicova16} find evidence for a similar
orbital variation to that observed here in the ionised phase of the absorption seen
during hard state. Finally, regardless of whether the orbital inclination
($\sim$27\deg) or the inner disk inclination ($\sim$40\deg) should be considered in
this regard, disk winds are not seen in X-ray binary systems viewed at similarly low
inclinations, even when they are in the soft state (\citealt{Ponti12}).

Lastly, we also see some level of variability in the flux of the narrow core of the iron
emission. This also shows some evidence for an orbital modulation (see Figure
\ref{fig_orbvar}, right panel), although this is much more marginal than for the
ionised absorption, being driven solely by epoch 4. Nevertheless, the evolution is
broadly consistent with the modulation expected should the narrow core arise
through reprocessing of the X-ray emission within the stellar wind, or even on the
stellar surface facing the black hole, as away from superior conjunection the body
of the stellar companion would block at least some of the line emitting region in both
cases.

From our observation closest to $\phi_{\rm{orb}} = 0.5$, we measure an equivalent
width of $10\pm5$\,eV for the narrow emission (see Table \ref{tab_param}). For
neutral iron, and the illuminating spectrum observed ($\Gamma \sim 2.6$),
reprocessing by material with solar abundances subtending a solid angle of 2$\pi$
should give an equivalent width of $\sim$100\,eV (\citealt{George91}). Correcting
for the projected area of the companion, given the stellar radius of $\sim$16.5\,\rsun,
the orbital separation of $\sim$40\,\rsun, and the orbital inclination of $\sim$27\deg\
(\citealt{Hanke09, Orosz11}), as well as the super-solar abundance inferred here,
we estimate an equivalent width of $\sim$20\,eV would be expected from the stellar
surface, if uniformly illuminated. This is slightly larger than observed. However, we
stress that this should be considered an upper limit, as the accretion disk around
\cyg\ must provide some shadowing of the stellar surface. Given the warp in the disk
inferred here, this effect is not trivial to estimate. Furthermore, some line emission
from the stellar wind from which \cyg\ accretes must also be present.
\cite{Torrejon10} find that the narrow line emission from HMXBs taken as an
ensemble is consistent with a roughly sperically distributed reprocessor around the
X-ray source, as expected for the stellar wind. \cite{Watanabe06} find that the iron
emission from the neutron star HMXB Vela X-1 requires contributions from both the
stellar wind and the stellar photosphere. It is possible that the narrow line emission
observed from \cyg\ is a similar blend of the two, but we are not able to make any
firm separation of their contributions based on the data considered here.

\section{Conclusions}
\label{sec_conc}

We have undertaken a multi-epoch analysis of soft-state observations of \cyg\ with
\nustar\ in order to investigate the spectral variability observed during this state.
Our detailed modeling, utilizing self-consistent reflection models to account for
reprocessing of the primary X-ray emission by the accretion disk, finds excellent
consistency across all epochs for the black hole spin, and the iron abundance of
the disk, quantities which should not vary on observational timescales. We confirm
that \cyg\ hosts a rapidly rotating black hole, finding $0.93 \lesssim a^{*} \lesssim
0.96$. This is in broad agreement with the majority of prior studies of the relativistic
disk reflection and also with constraints on the spin obtained through studies of the
thermal accretion disk continuum. The iron abundance obtained is super-solar
(Fe/solar $\sim$ 4), in qualitative agreement with previous studies.

Our work also confirms the apparent misalignment between the inner disk and the
orbital plane of the binary system reported previously. We find that the magnitude
of this warp is $\sim$10--15\deg\ ($i_{\rm{orb}} \sim 27$\deg, and we find
$i_{\rm{disk}} \sim 40$\deg), and that the level of misalignment does not appear to
vary significantly with time or accretion state. This does not significantly change,
and may even improve, the agreement between the reflection results presented
here and the thermal continuum results regarding the black hole spin.

The spectral variability observed by \nustar\ is dominated by variations in the
primary continuum. We find evidence that the temperature of the scattering
electron plasma changes from epoch to epoch, causing the variations observed in
the spectrum at the highest energies probed by \nustar. Finally, all epochs show
absorption from ionized iron at $\sim$6.7\,keV. The strength of this absorption
varies across the orbital phase of \cyg\ in a manner consistent with the absorbing
material being an ionized phase of the focused stellar wind from its supergiant
companion.

\section*{ACKNOWLEDGEMENTS}

The authors would like to thank the referee for their prompt and useful feedback.
DB acknowledges financial support from the French Space Agency (CNES).
VG acknowledges financial support provided by NASA through the Smithsonian
Astrophysical Observatory (SAO) contract SV3-73016 to MIT for Support of the
\chandra\ X-Ray Center (CXC) and Science Instruments; CXC is operated by SAO
for and on behalf of NASA under contract NAS8-03060. This research has made
use of data obtained with \nustar, a project led by Caltech, funded by NASA and
managed by NASA/JPL, and has utilized the \nustardas\ software package, jointly
developed by the ASI Science Data Center (ASDC, Italy) and Caltech (USA).
\swift\ BAT transient monitor results are provided by the \swift\ BAT team. This
research has also made use of \maxi\ data provided by RIKEN, JAXA and the
\maxi\ team.

{\it Facilites:} \facility{NuSTAR}

%\pagebreak
\appendix

\section{\textit{NuSTAR} Spacecraft Science (Mode 6) Data Reduction}
\label{app_mode6}

\begin{figure}[!t]
\hspace*{-0.35cm}
\epsscale{1.16}
\plotone{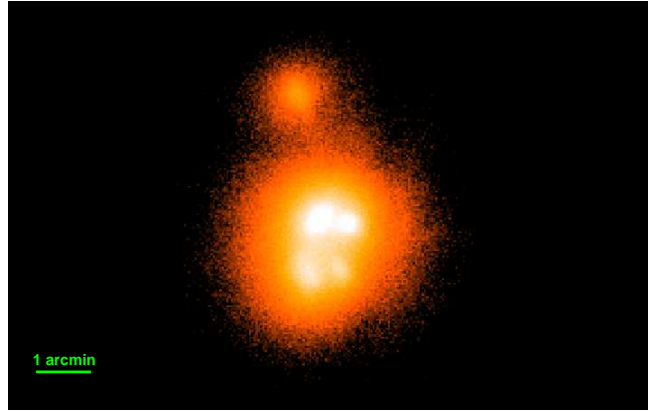}
\caption{
Sky image extracted from the full FPMA spacecraft science (mode 6) event file for one
of the \nustar\ observations of \cyg\ (OBSID 10002003001). Owing to the switching
between different CHU combinations, multiple centroids from the same source can be
observed. In this extreme case, 5 centroids are seen.
}
\vspace{0.2cm}
\label{fig_mode6_img}
\end{figure}

\begin{figure*}
\hspace*{-0.7cm}
\epsscale{0.53}
\plotone{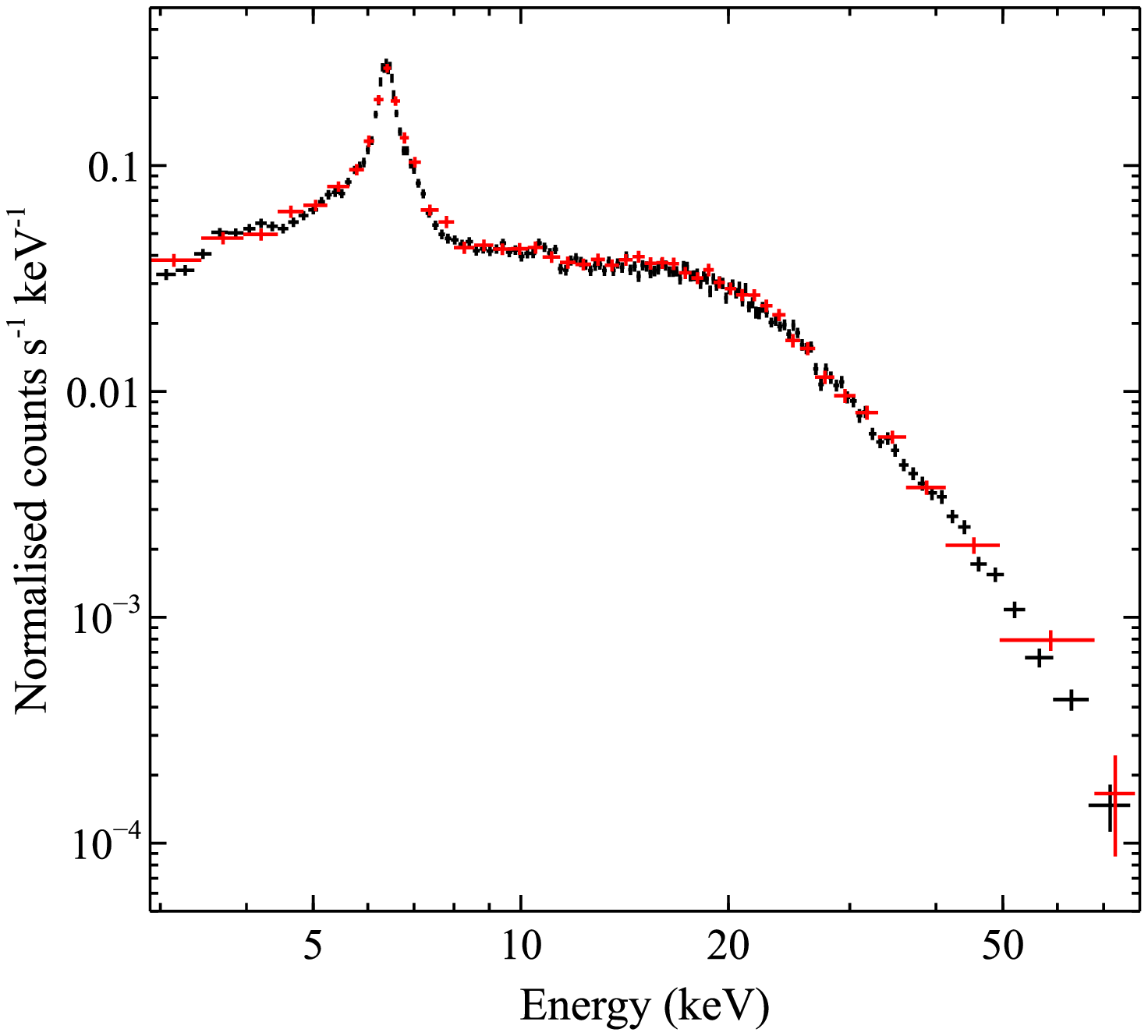}
\hspace*{0.9cm}
\plotone{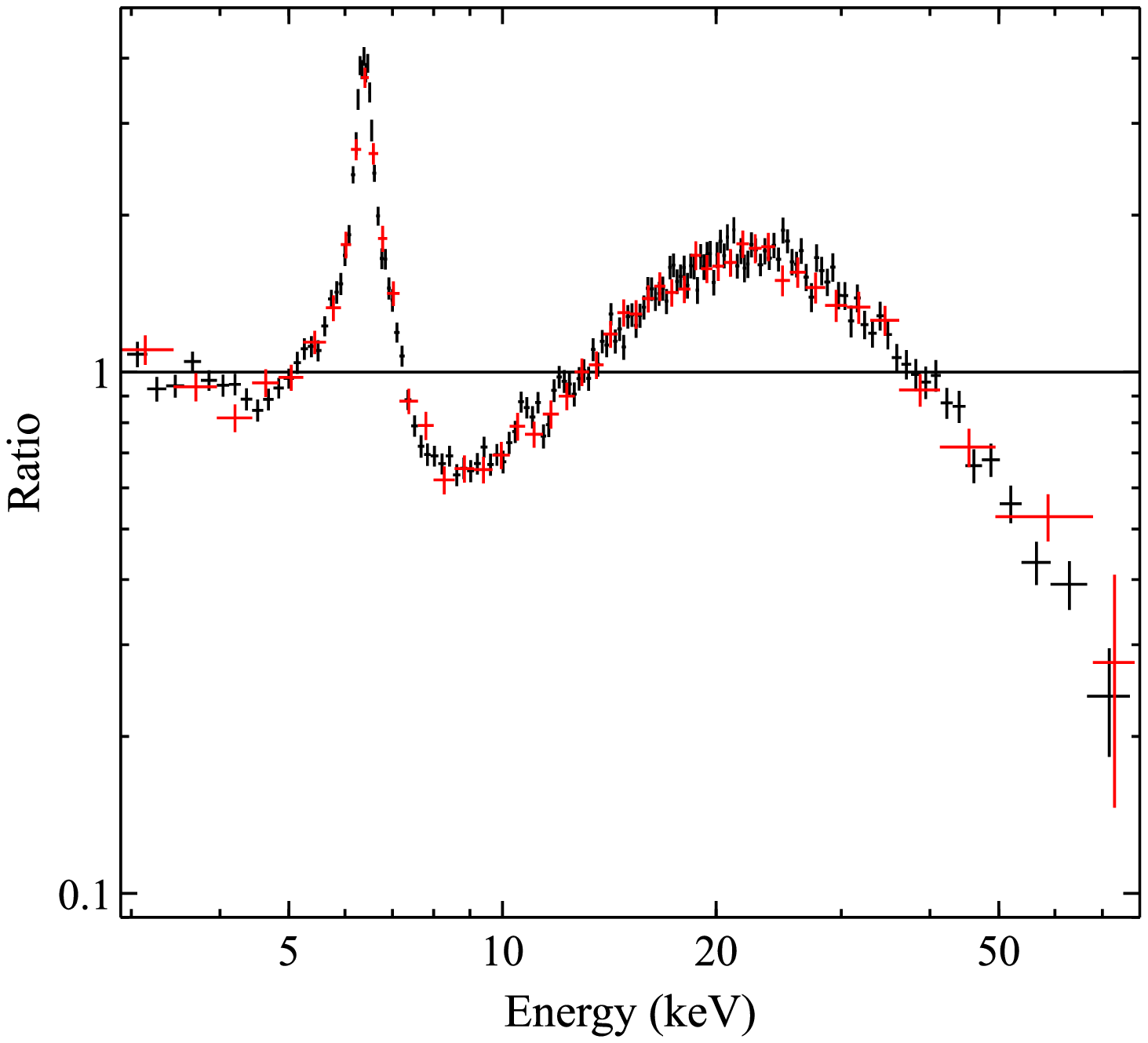}
\caption{
Spectra extracted from the targeted \nustar\ observation of the Circinus galaxy
nucleus (OBSID 60002039002). The regular science data (mode 1) are shown in
black, and the spacecraft science data (mode 6) are shown in red; only FPMA
data are shown for clarity. We plot both the count spectrum (\textit{left panel}) and
a data/model ratio to a powerlaw model (\textit{right panel}). Excellent agreement
is seen between the two spectra, as expected given the lack of variability
observed from this source
(\citealt{Arevalo14}).
}
\vspace{0.2cm}
\label{fig_mode6_circinus}
\end{figure*}

For the \nustar\ mission, ``spacecraft science'' (mode 6) events refer to those
collected during periods in which an aspect solution is not available from the
on-board star tracker located on the X-ray optics bench (Camera Head Unit 4,
or CHU4, which is the primary method for determining the absolute pointing;
see \citealt{NUSTAR}). In this situation when CHU4 is not available, because
it is either blinded by a bright target or Earth-occultation, the aspect
reconstruction (source sky coordinate calculation) is calculated using the
spacecraft bus attitude solution. In normal operations when the aspect
reconstruction uses CHU4 to calculate source coordinates, the accuracy is
$\pm$8$''$. Using the spacecraft bus in mode 6, this error increases to $\sim
$2$'$.

The inaccuracies incurred from the spacecraft bus attitude solution come about
due to thermal flexure in the mounting of the spacecraft bus star cameras (CHU1,
2, 3), which are unique for each pointing and cannot be modeled. They manifest
themselves in the calculated source coordinates, which can cause the sky image
of the source to appear with multiple centroids as shown in Figure
\ref{fig_mode6_img}. There are a total of three star tracker CHUs on the
spacecraft bus with a total of 7 different combinations. Each one of those
combinations will have a unique offset and a typical observation has between 2
to 5 of such CHU combinations. The severity of the offsets is dependent on the
Solar aspect angle and some unknown variables that make them unpredictable.
As such, mode 6 data is \textit{not} recommended for applications in which
imaging capability is necessary.

Since all responses are calculated in the optics frame, whose relation to the
detector plane is accurately tracked by a laser system, the spectrum of the
source remains unaffected by mode 6. The only problem with using mode 6 is
the degradation of the PSF and the challenges with choosing the correct spectral
extraction regions, since the optimal region for one CHU combination will not be
the optimal region for another CHU combination. Effective areas will be calculated
for the center of a region and thus if a centroid is far outside the region it will be
assigned the wrong effective area.

For sources with concentrated centroids (differences of $<$ 1$'$) simply choosing
a large region is sufficient. If the source centroids are scattered (difference $>$
1$'$) then it will be necessary to divide the observation into periods that correspond
to the individual CHU combinations that were present during the observation and
extract the spectra from these periods separately. This can be done by filtering on
the housekeeping file \texttt{hk/xx\_chu123.fits} for the different CHU combinations.
The file contains the attitude solutions from the three spacecraft bus star trackers
CHU1, CHU2, and CHU3. There is an extension for each CHU and the
\texttt{VALID} column marks whether the CHU was active (1 = on, 0 = off). In
addition, the columns \texttt{RESIDUAL}, \texttt{STARSFAIL}, \texttt{OBJECTS},
and \texttt{QCHU[3]} should be included in the filtering. A particular CHU was active
if:
\begin{itemize}
\item \texttt{VALID} = 1
\item \texttt{RESIDUAL} $<$ 20 %(arcseconds)
\item \texttt{STARSFAIL} $<$ \texttt{OBJECTS}
\item \texttt{QCHU[3]} $\neq$ 1
\end{itemize}

Here, \texttt{RESIDUAL} is a residual of the aspect solution fit, \texttt{OBJECTS} is
the number of point sources detected by the star trackers, \texttt{STARSFAIL} is the
number of these objects without known stellar counterparts in the reference catalogue
used (which is based on the Hipparcos and Tycho-2 catalogues; \citealt{Hipparcos,
Tycho2}) and \texttt{QCHU[3]} is the real part of the position quaternion solution for
the spacecraft orientation, which must be \texttt{QCHU[3]} $\neq$ 1 for a valid solution.
As mentioned above, there are 7 possible combinations: CHU1 only, CHU2 only,
CHU3 only, CHU1\&2, CHU1\&3, CHU2\&3, and CHU1\&2\&3. Generating GTI’s for
each combination will separate the centroids in time. There may sometimes still be
weak centroids remaining from other CHU combinations that come from spurious 
solutions interspaced with the primary CHU combination the data are filtered on,
caused in part because not all solutions are telemetered to the ground, and switches
between CHU combinations could happen in between. Should this prove to be the
case, source regions should be selected to include all the counts from each of these
centroids such that the PSF correction applied by the pipeline will be as accurate as
possible.

After spectra have been extracted for each CHU combination, they can be
combined as standard with tools like \addascaspec. ARFs must be generated and
combined for each CHU combination, but new RMFs do not necessarily need to be
generated if all the events fall on the same detector as the regular (mode 1) data. 

In order to demonstrate the reliability of the mode 6 data, we show in Figure
\ref{fig_mode6_circinus} a comparison of the spectra extracted from modes 1 and
6 for the nucleus of the Circinus galaxy. This is a bright source that is unresolved
by \nustar, and as a Compton-thick AGN is known to exhibit a stable flux. We
therefore expect good agreement between the two extractions. A full scientific
analysis of this source is presented in \cite{Arevalo14}. For brevity we use only the
data from OBSID 60002039002, in which the nucleus was on-axis as the primary
target; we do not consider the additional \nustar\ observations of the Circinus
galaxy which were centred on the nearby ultraluminous X-ray source (ULX;
\citealt{Walton13culx}).

The mode 1 data are reduced following standard procedure (see section \ref{sec_red}),
with source spectra extracted from a circular region of radius 100$''$, and the 
background was estimated from a nearby region on the same detector, avoiding the
ULX. The mode 6 data was reduced following the procedure outlined above. Two
centroids can be seen in the mode 6 image, but one dominates and the offset of the
other is not very large, so we extract source spectra from a single, slightly larger
circular region of radius of radius 110$''$ which incorporates both; background was
estimated in the same manner as for mode 1. The good mode 1 exposure for this
observation is $\sim$54\,ks, and the good mode 6 exposure is $\sim$12\,ks. As can be
seen from Figure \ref{fig_mode6_circinus}, the two spectra of the nucleus do indeed
agree well, and the fluxes returned for the two modes are consistent to within 2\,\% for
both FPMA and FPMB.

\bibliographystyle{/Users/dwalton/papers/apj}

\bibliography{/Users/dwalton/papers/references}

\label{lastpage}

\end{document}